\documentclass[traditabstract]{aa} 
\usepackage{txfonts}
\usepackage{graphicx}
\usepackage{longtable}

\begin{document}

\title{Massive unseen companions to hot faint underluminous stars from SDSS (MUCHFUSS)}
\subtitle{
Analysis of seven close subdwarf B binaries
\thanks{Based on observations at the Paranal Observatory of the European 
Southern Observatory for programme number 081.D-0819. Based on observations at the La Silla Observatory of the 
European Southern Observatory for programmes number 082.D-0649 and 084.D-0348. Based on observations collected at 
  the Centro Astron\'omico Hispano Alem\'an (CAHA) at Calar Alto, operated 
  jointly by the Max-Planck Institut f\"ur Astronomie and the Instituto de
   Astrof\'isica de Andaluc\'ia (CSIC). Based on observations with the William Herschel Telescope and the Isaac Newton Telescope operated both by the Isaac Newton Group at the Observatorio del Roque de los Muchachos of the Instituto de Astrofisica de Canarias on the island of La Palma, Spain. Based on observations with the Southern Astrophysical Research (SOAR) telescope operated by the U.S. National Optical Astronomy Observatory (NOAO), the Ministério da Ciencia e Tecnologia of the Federal Republic of Brazil (MCT), the University of North Carolina at Chapel Hill (UNC), and Michigan State University (MSU). Based on observations obtained at the Gemini Observatory, which is operated by the
Association of Universities for Research in Astronomy, Inc., under a cooperative agreement
with the NSF on behalf of the Gemini partnership: the National Science Foundation (United
States), the Science and Technology Facilities Council (United Kingdom), the
National Research Council (Canada), CONICYT (Chile), the Australian Research Council
(Australia), Ministério da Ciência e Tecnologia (Brazil) 
and Ministerio de Ciencia, Tecnología e Innovación Productiva  (Argentina). This paper uses observations made at the South African Astronomical Observatory (SAAO).}
}

\author{S. Geier \inst{1}
   \and P. F. L. Maxted \inst{2}
   \and R. Napiwotzki \inst{3}
   \and R. H. \O stensen \inst{4}
   \and U. Heber \inst{1}
   \and H. Hirsch \inst{1}
   \and T. Kupfer \inst{1}
   \and S. M\"uller \inst{1}
   \and A. Tillich \inst{1}
   \and B. N. Barlow \inst{5}
   \and R. Oreiro \inst{4,6}
   \and T. A. Ottosen \inst{7,8}
   \and C. Copperwheat \inst{9}
   \and B. T. G\"ansicke \inst{9}
   \and T. R. Marsh \inst{9}}

\offprints{S.\,Geier,\\ \email{geier@sternwarte.uni-erlangen.de}}

\institute{Dr. Karl Remeis-Observatory \& ECAP, Astronomical Institute,
Friedrich-Alexander University Erlangen-Nuremberg, Sternwartstr. 7, D 96049 Bamberg, Germany
\and Astrophysics Group, Keele University, Staffordshire, ST5 5BG, UK
\and Centre of Astrophysics Research, University of Hertfordshire, College
  Lane, Hatfield AL10 9AB, UK
\and Institute of Astronomy, K.U.Leuven, Celestijnenlaan 200D, B-3001 Heverlee, Belgium
\and Department of Physics and Astronomy, University of North Carolina, Chapel Hill, NC 27599-3255, USA
\and Instituto de Astrof\'\i sica de Andaluc\'\i a Glorieta de la Astronom\'\i a s/n 18008 Granada, Spain
\and Department of Physics and Astronomy, Aarhus University, 8000 Aarhus C, Denmark
\and Nordic Optical Telescope, Apartado 474, E-38700 Santa Cruz de La Palma, Santa Cruz de Tenerife, Spain
\and Department of Physics, University of Warwick, Conventry CV4 7AL, UK
}

\date{Received \ Accepted}

\abstract{
The project Massive Unseen Companions to Hot Faint Underluminous Stars from SDSS (MUCHFUSS) aims at finding hot subdwarf stars with massive compact companions like massive white dwarfs ($M>1.0\,{\rm M_{\odot}}$), neutron stars or stellar mass black holes. The existence of such systems is predicted by binary evolution theory and recent discoveries indicate that they exist in our Galaxy.\\ 
First results are presented for seven close binary sdBs with short orbital periods ranging from $\simeq0.21\,{\rm d}$ to $1.5\,{\rm d}$. The atmospheric parameters of all objects are compatible with core helium-burning stars. The companions are most likely white dwarfs. In one case the companion could be shown to be a white dwarf by the absence of light-curve variations. However, in most cases late type main sequence stars cannot be firmly excluded. Comparing our small sample with the known population of close sdB binaries we show that our target selection method aiming at massive companions is efficient. The minimum companion masses of all binaries in our sample are high compared to the reference sample of known sdB binaries.\\ 

}

\maketitle

\section{Introduction \label{s:intro}}

Subluminous B stars (sdBs) are core helium-burning stars with very thin hydrogen envelopes and masses around $0.5M_{\rm \odot}$ 
(Heber \cite{heber86}, see Heber \cite{heber09} for a review). A large fraction of the sdB stars ($40\,\%$ to $80\,\%$) are members of short period binaries (Maxted et al. \cite{maxted01};  Napiwotzki et al. \cite{napiwotzki04a}). Several studies were undertaken to determine the orbital parameters of sub\-dwarf binaries, and found periods ranging from $0.07$ to more than $10\,{\rm d}$ with a peak at $0.5$ to $1.0\,{\rm d}$ (e.g. Edelmann et al. \cite{edelmann05}; Morales-Rueda et al. \cite{morales03}). For close binary sdBs, common envelope (CE) ejection is the most probable formation channel. In this scenario two main sequence stars of different masses evolve in a binary system. The heavier one will reach the red giant phase first and fill its Roche lobe. If the mass transfer to the com\-panion is dynamically unstable, a common envelope is formed. Due to friction the two stellar cores lose orbital energy, which is deposited within the envelope and leads to a shortening of the binary period. Eventually the common envelope is ejected and a close binary system is formed, which contains a core helium-burning sdB and a main sequence companion. If the companion has already evolved to a white dwarf (WD) when the red giant fills its Roche lobe, a close sdB+WD binary is formed (Han et al. \cite{han02,han03}). Under certain conditions, two consecutive CE phases are possible as well. 

\begin{table}[t!]
\caption{Solved binary systems.} 
\label{tab:solved}
\begin{center}
\begin{tabular}{llll} 
SDSS name & short name & other names \\
& & \\
\hline
\\[-3mm]
SDSS\,J002323.99$-$002953.2 & J0023$-$0029 & PB\,5916 \\
SDSS\,J113840.68$-$003531.7 & J1138$-$0035 & PG\,1136-003 \\
SDSS\,J150513.52+110836.6 & J1505+1108 & PG\,1502+113 \\
SDSS\,J165404.25+303701.7 & J1654+3037 & PG\,1652+307 \\
SDSS\,J172624.09+274419.3 & J1726+2744 & PG\,1724+278 \\
SDSS\,J204613.40$-$045418.7 & J2046$-$0454 & $-$ \\
SDSS\,J225638.34+065651.0 & J2256+0656 & PG\,2254+067 \\
\hline \\[-3mm]
\end{tabular}
\end{center}
\end{table}

In general it is difficult to put constraints on the nature of the close companions to sdB stars. Since most of the binaries are single-lined, only lower limits have been derived from the binary mass functions, which are in general compatible with main sequence stars of spectral type M or compact objects like white dwarfs. Only in special and hence rare cases can tighter constraints be put on the nature of the companions. 

Subdwarf binaries with massive WD companions turned out to be candidates for supernova type Ia (SN Ia) progenitors because these systems lose angular momentum due to the emission of gravitational waves and start mass transfer. This mass transfer, either  from accretion of He onto the WD during the sdB phase (e.g. Yoon \& Langer \cite{yoon04} and references therein), or the subsequent merger of the system after the sdB star itself has turned into a WD (Tutukov \& Yungelson \cite{tutukov81}; Webbink \cite{webbink84}) may cause the companion to approach the Chandrasekhar limit and explode as SN Ia.

SN~Ia play a key role in the study of cosmic evolution (e.g. Riess et al. \cite{riess98}; Leibundgut \cite{leibundgut01}; Perlmutter et al. \cite{perlmutter99}). One of the best known candidate systems for the double degenerate merger scenario is the sdB+WD binary KPD\,1930$+$2752 (Maxted et al. \cite{maxted00}; Geier et al. \cite{geier07}). Mereghetti et al. (\cite{mereghetti09}) showed that in the X-ray binary HD\,49798 a massive ($>1.2\,M_{\rm \odot}$) white dwarf accretes matter from a closely orbiting subdwarf O companion. The predicted amount of accreted material is sufficient for the WD to reach the Chandrasekhar limit. This makes HD\,49798 another candidate for SN\,Ia progenitor. Furthermore,  Perets et al. (\cite{perets10}) showed that helium accretion onto a white dwarf may be responsible for a subclass of faint and calcium-rich SN Ib events.

Geier et al. (\cite{geier08}, \cite{geier10a}, \cite{geier10b}) analysed high resolution spectra of sdB stars in close binaries. Assuming synchronised rotation they constrained the masses and the nature of the unseen companions in 31 cases. While most of the derived companion masses were consistent with either late type main sequence stars or white dwarfs, the compact companions of some sdBs may be either massive white dwarfs, neutron stars (NS) or stellar mass black holes (BH). However, Geier et al. (\cite{geier10b}) also showed that the assumption of orbital synchronisation in close sdB binaries is not always justified and that their analysis suffers from huge selection effects. 

The existence of sdB+NS/BH systems is predicted by binary evolution theory (Podsiadlowski et al. \cite{podsi02}; Pfahl et al. \cite{pfahl03}). The formation channel includes two phases of unstable mass transfer and one supernova explosion. The fraction of sdB+NS/BH systems is predicted to be about $2\%$ of the close sdB binaries (Geier et al. \cite{geier10b}). Yungelson \& Tutukov (\cite{yungelson05}) and Nelemans (\cite{nelemans10}) performed independent binary evolution calculations and confirm that sdB+NS/BH systems should exist. According to the results of Nelemans (\cite{nelemans10}) about $1\%$ of the subdwarfs in close binaries should have a neutron star companion, whereas only $0.01\%$ should be orbited by a black hole. Yungelson \& Tutukov (\cite{yungelson05}) predict the sdB+NS fraction to be of the order of $0.8\%$.

Since sdB stars eventually evolve to WDs there should also exist a population of white dwarfs with massive compact companions. Badenes et al. (\cite{badenes09}) reported the discovery of a close binary consisting of a massive white dwarf and an unseen neutron star or black hole companion, but Marsh et al. (\cite{marsh10}) most recently showed that the system is double-lined and consists of a massive white dwarf orbited by a low mass white dwarf. The system mass is below the Chandrasekhar limit. Their results were confirmed by Kulkarni \& van Kerkwijk (\cite{kulkarni10}). Common envelope ejection was proposed as the most likely formation channel for the binary PSR\,J1802$-$2124, which consists of a millisecond pulsar and a CO white dwarf in close orbit ($P=0.7\,{\rm d}$, Ferdman et al. \cite{ferdman10}). This peculiar system may have evolved through an earlier sdB+NS phase.

\section{The MUCHFUSS project}\label{s:much}

The discovery of sdB binary candidates with massive compact companions provides a first hint that a whole population of non-interacting binaries with such companions may be present in our Galaxy. The known candidate sdB+NS/BH binaries have low orbital inclinations ($15-30^{\rm \circ}$, Geier et al. \cite{geier10b}). High inclination systems must exist as well and should be more numerous. In this case a determination of the orbital parameters is sufficient to put a lower limit to the companion mass by calculating the binary mass function. If this lower limit exceeds the Chandrasekhar mass and no sign of a companion is visible in the spectra, the existence of a massive compact companion is proven without the need for any additional assumptions. 

The project Massive Unseen Companions to Hot Faint Underluminous Stars from SDSS\footnote{Sloan Digital Sky Survey} (MUCHFUSS) aims at finding sdBs with compact companions like massive white dwarfs ($M>1.0\,{\rm M_{\odot}}$), neutron stars or black holes. About $80$ binaries have been selected for follow-up. Survey and target selection are described in detail in Geier et al. (\cite{geier10c}). The same selection criteria that we applied to find such binaries are also well suited to single out hot subdwarf stars with constant high radial velocities (RV) in the Galactic halo and search for hypervelocity stars. First results of this second part of the project (Hyper-MUCHFUSS) are presented in Tillich et al. (\cite{tillich10}). 

Here we present the spectroscopic analysis of the first sdB binaries discovered in the course of the MUCHFUSS project (see Table~\ref{tab:solved}). In Sect.~\ref{s:data} the observations and the data reduction are described. Sects.~\ref{s:orbit} and \ref{s:atmo} deal with the determination of the orbital and atmospheric parameters of the sdB stars. Sect.~\ref{s:comp} explains the way the minimum masses of the unseen companions are constrained, while results are presented in Sect.~\ref{s:results}. The efficiency of our target selection is discussed in Sect.~\ref{s:efficient}, a short summary and an outlook are eventually given in Sect.~\ref{s:summary}.

\begin{figure*}[t!]
\begin{center}
	\resizebox{17cm}{!}{\includegraphics{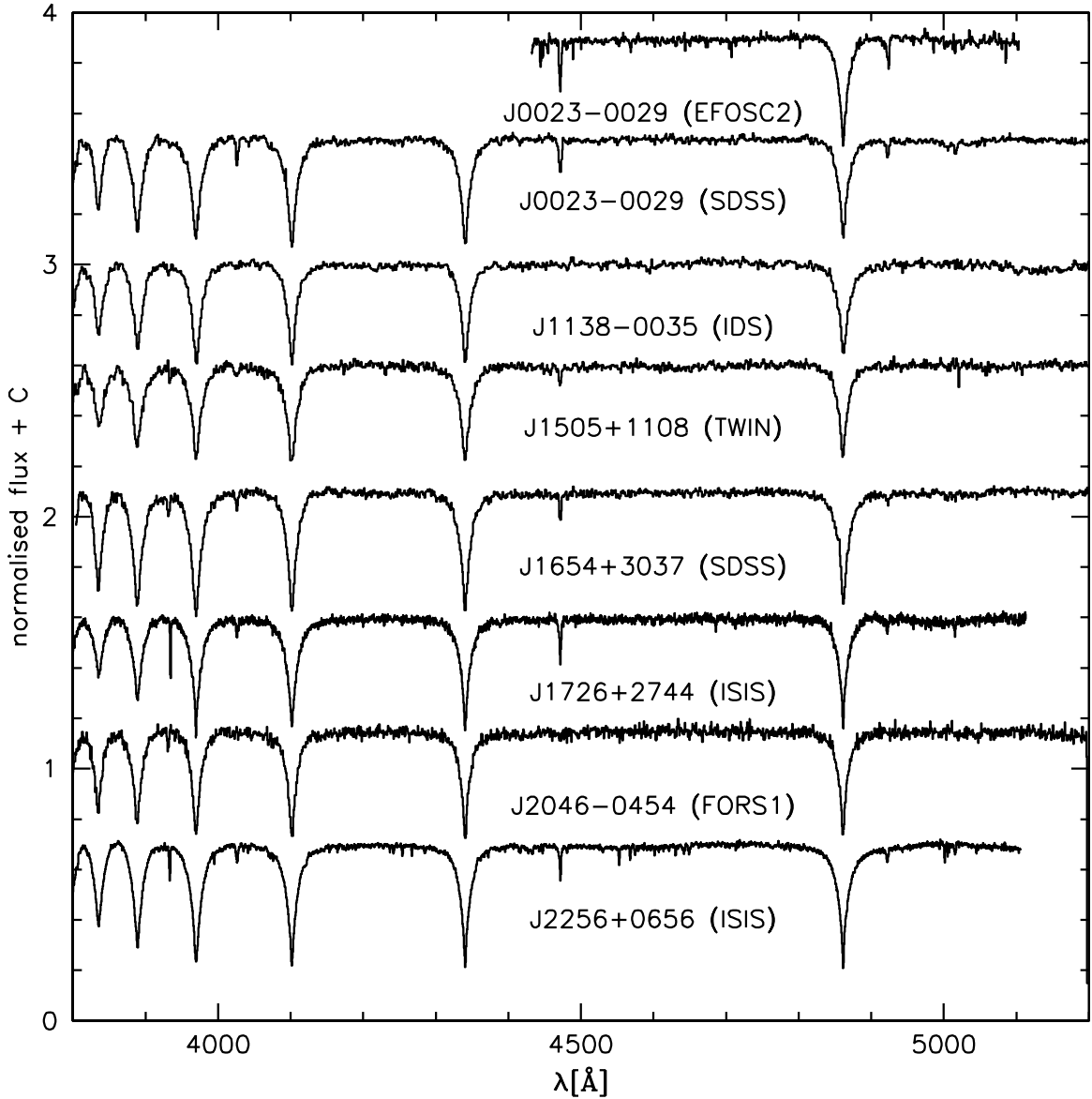}}
\end{center}
\caption{Medium resolution spectra of the programme stars taken with different instruments. Multiple observations of the same target  
have been shifted to rest wavelength and coadded.}
\label{specexample}
\end{figure*}

\begin{figure*}[t!]
\begin{center}
	\resizebox{8.5cm}{!}{\includegraphics{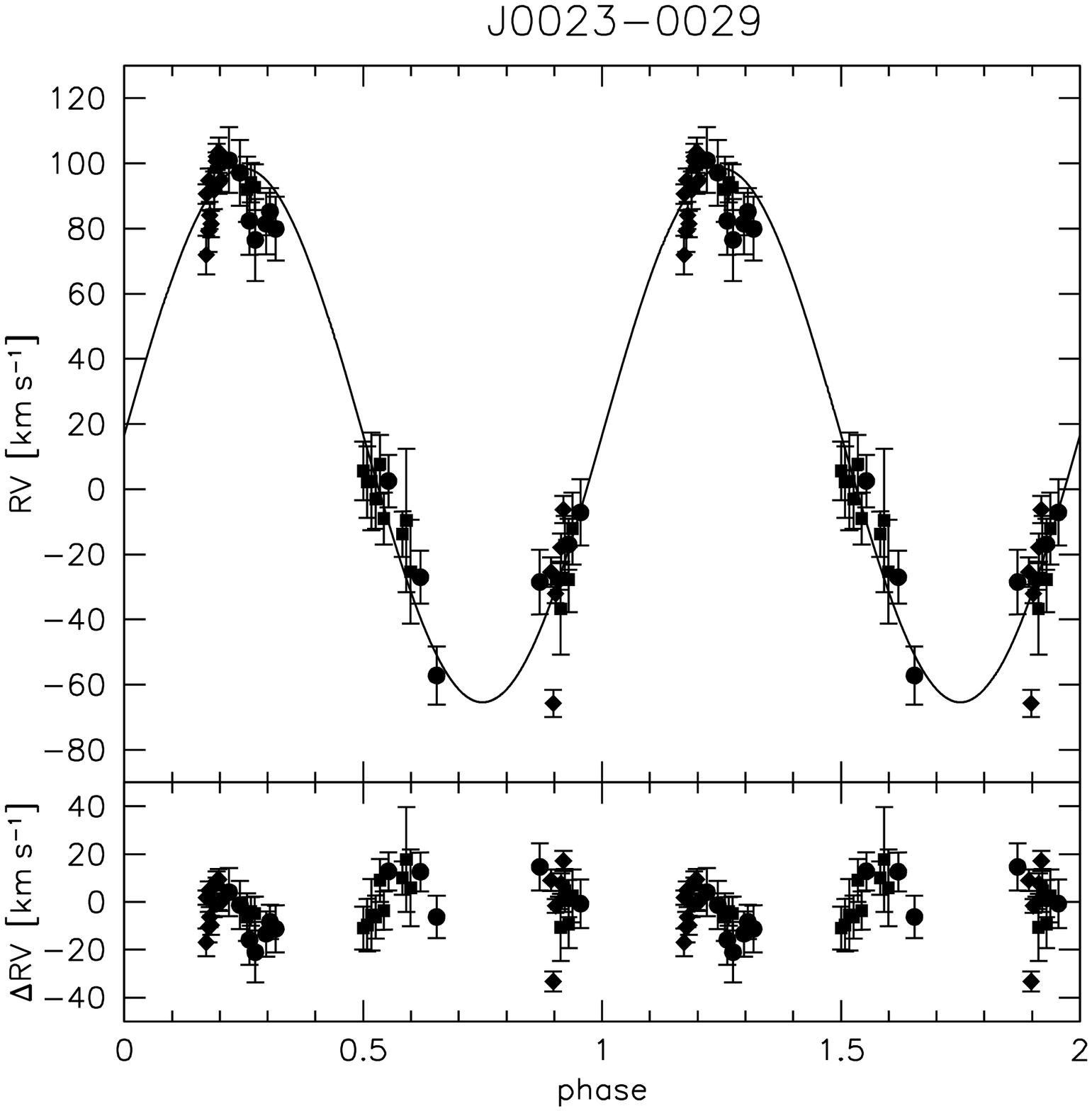}}
        \resizebox{8.5cm}{!}{\includegraphics{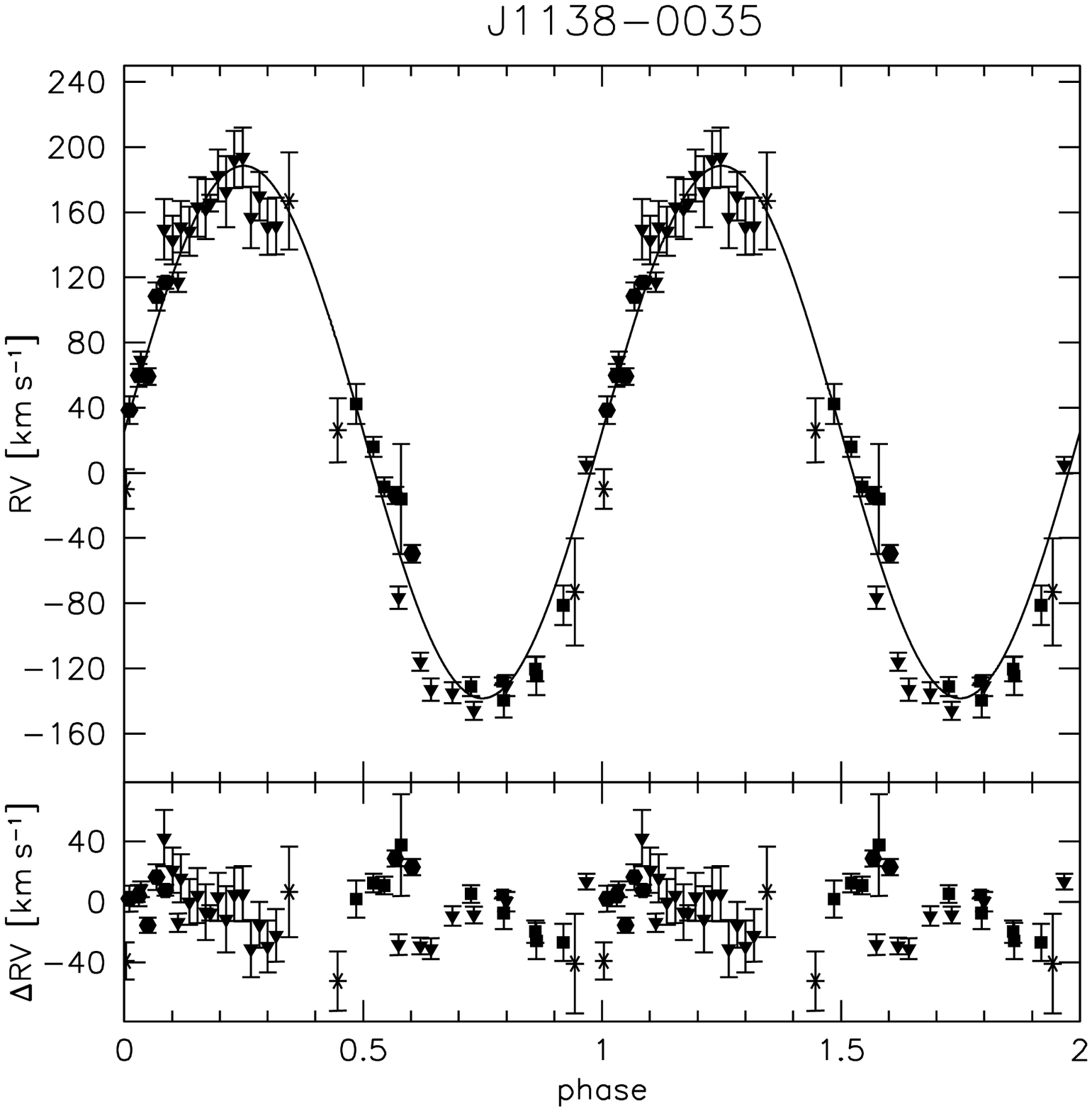}}
        \resizebox{8.5cm}{!}{\includegraphics{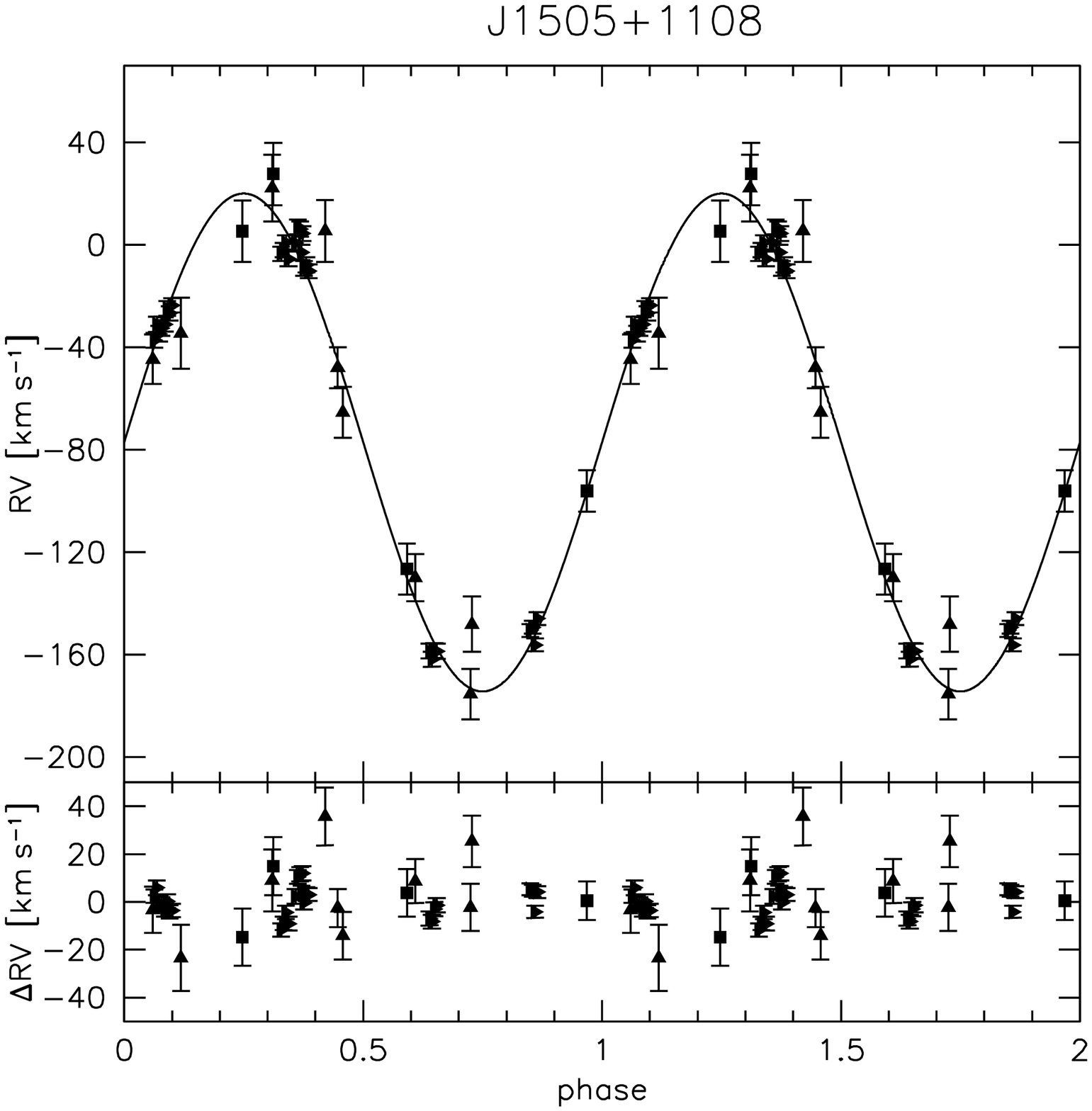}}
        \resizebox{8.5cm}{!}{\includegraphics{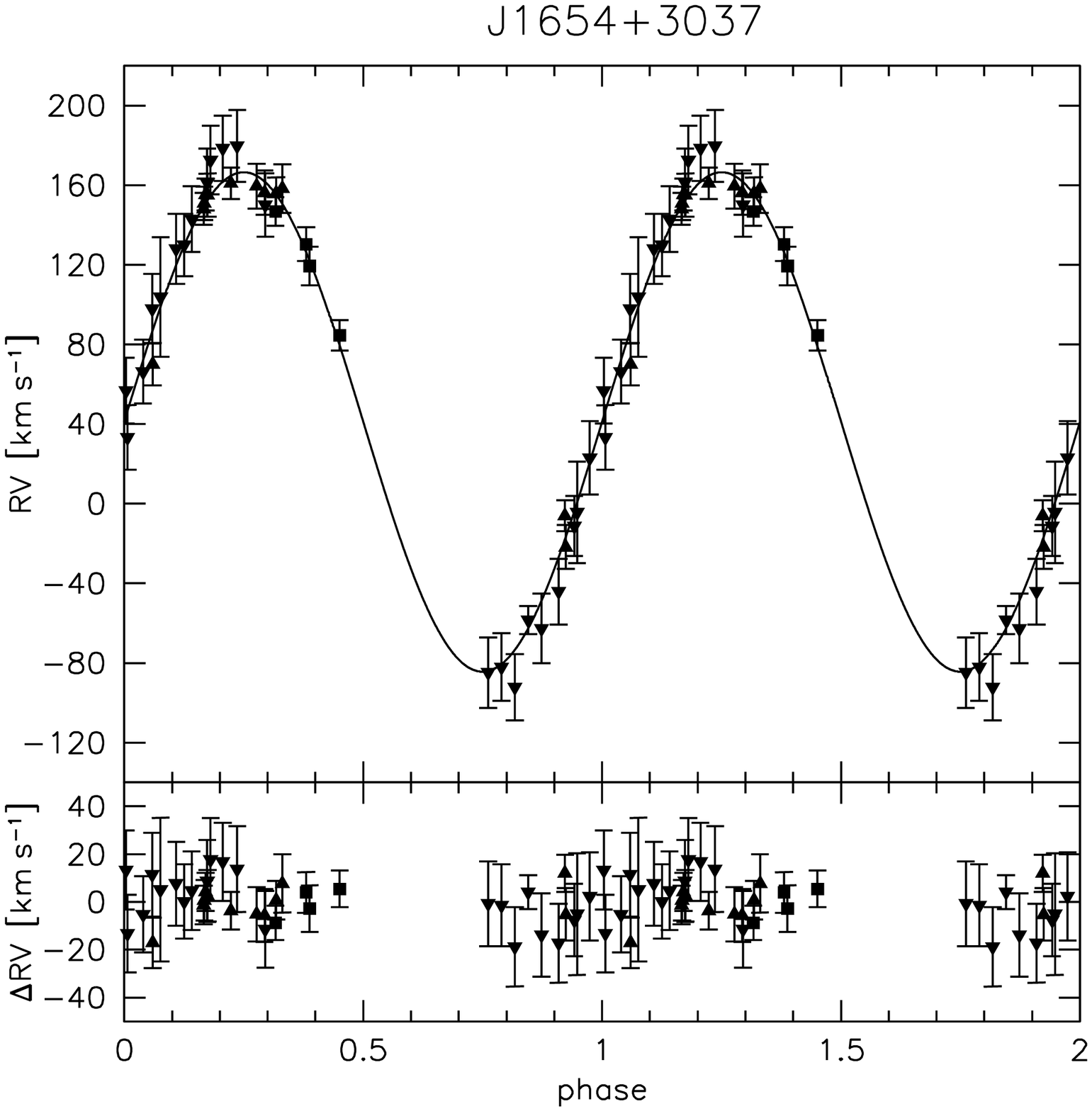}}
\end{center}
\caption{Radial velocity plotted against orbital phase. The RV data were phase folded with the most likely orbital periods. The residuals are plotted below. The RVs were measured from spectra obtained with SDSS (rectangles), CAHA3.5m/TWIN (upward triangles), WHT/ISIS (diamonds), INT/IDS (downward triangles), ESO-VLT/FORS1 (triangles turned to the left), Gemini/GMOS (triangles turned to the right), ESO-NTT/EFOSC2 (circles), SOAR/Goodman (hexagons) and SAAO-1.9m/Grating (stars).}
\label{rv1}
\end{figure*}
 
\begin{figure*}[t!]
\begin{center} 
        \resizebox{8.5cm}{!}{\includegraphics{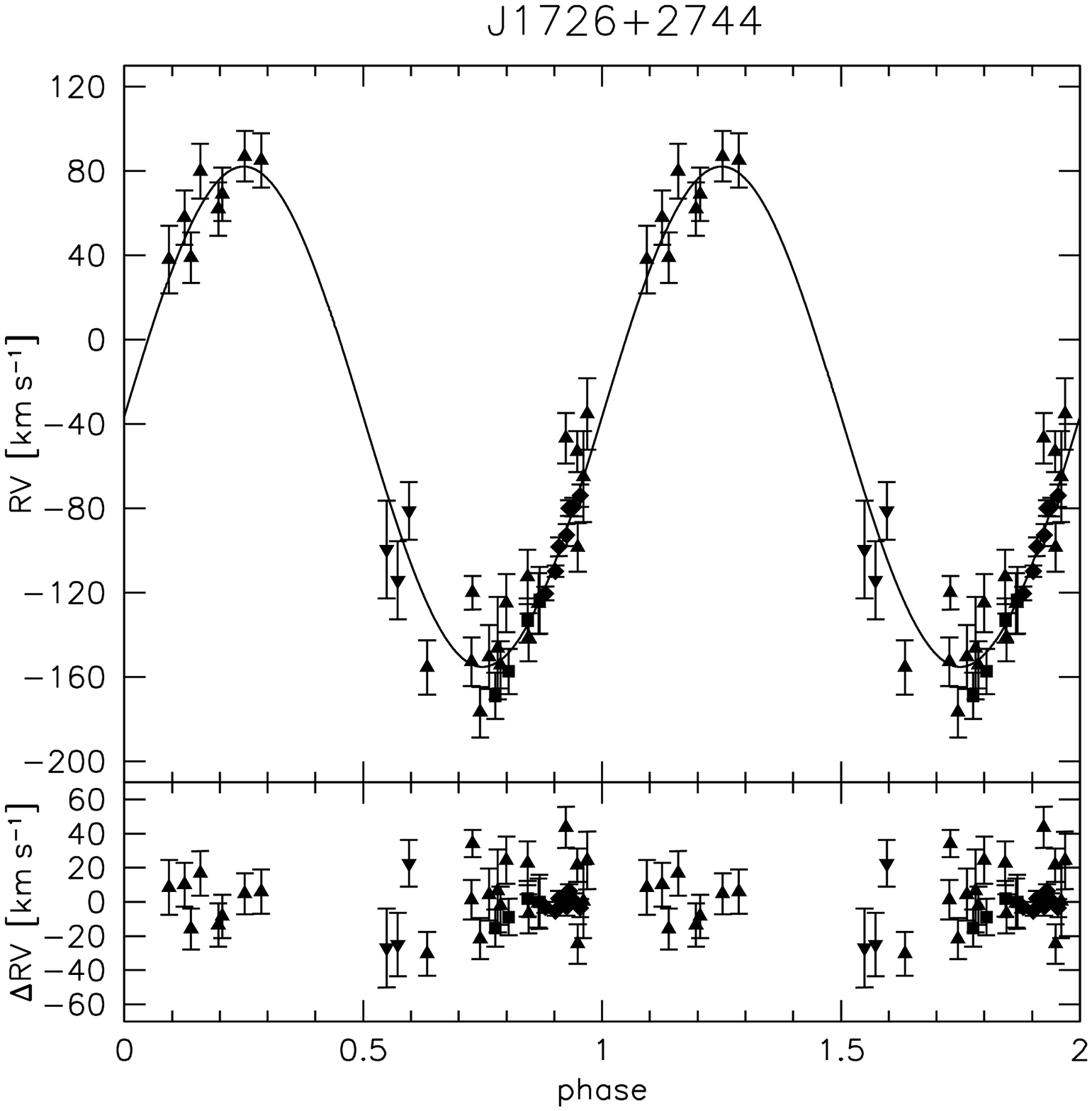}}
        \resizebox{8.5cm}{!}{\includegraphics{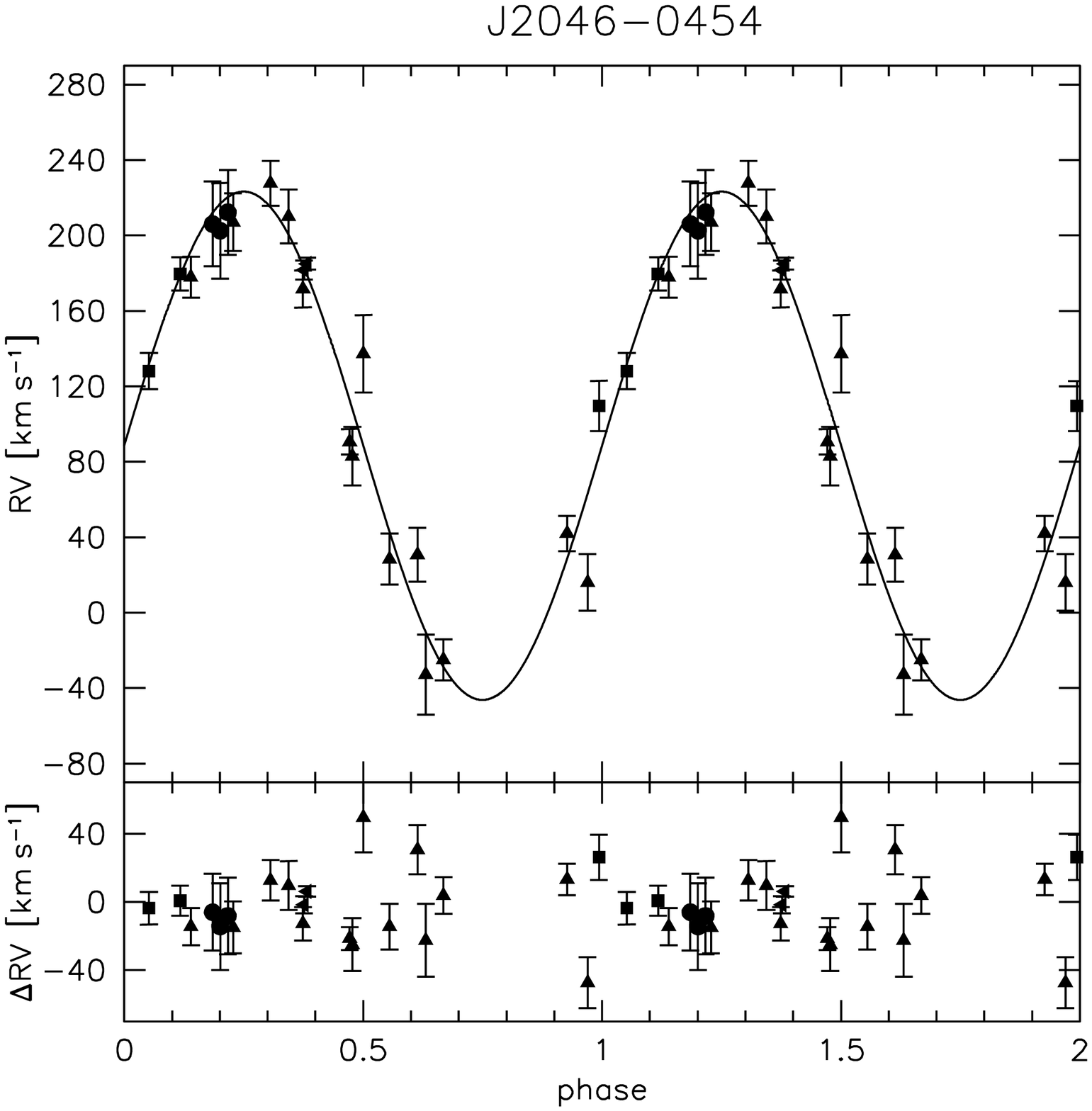}}
        \resizebox{8.5cm}{!}{\includegraphics{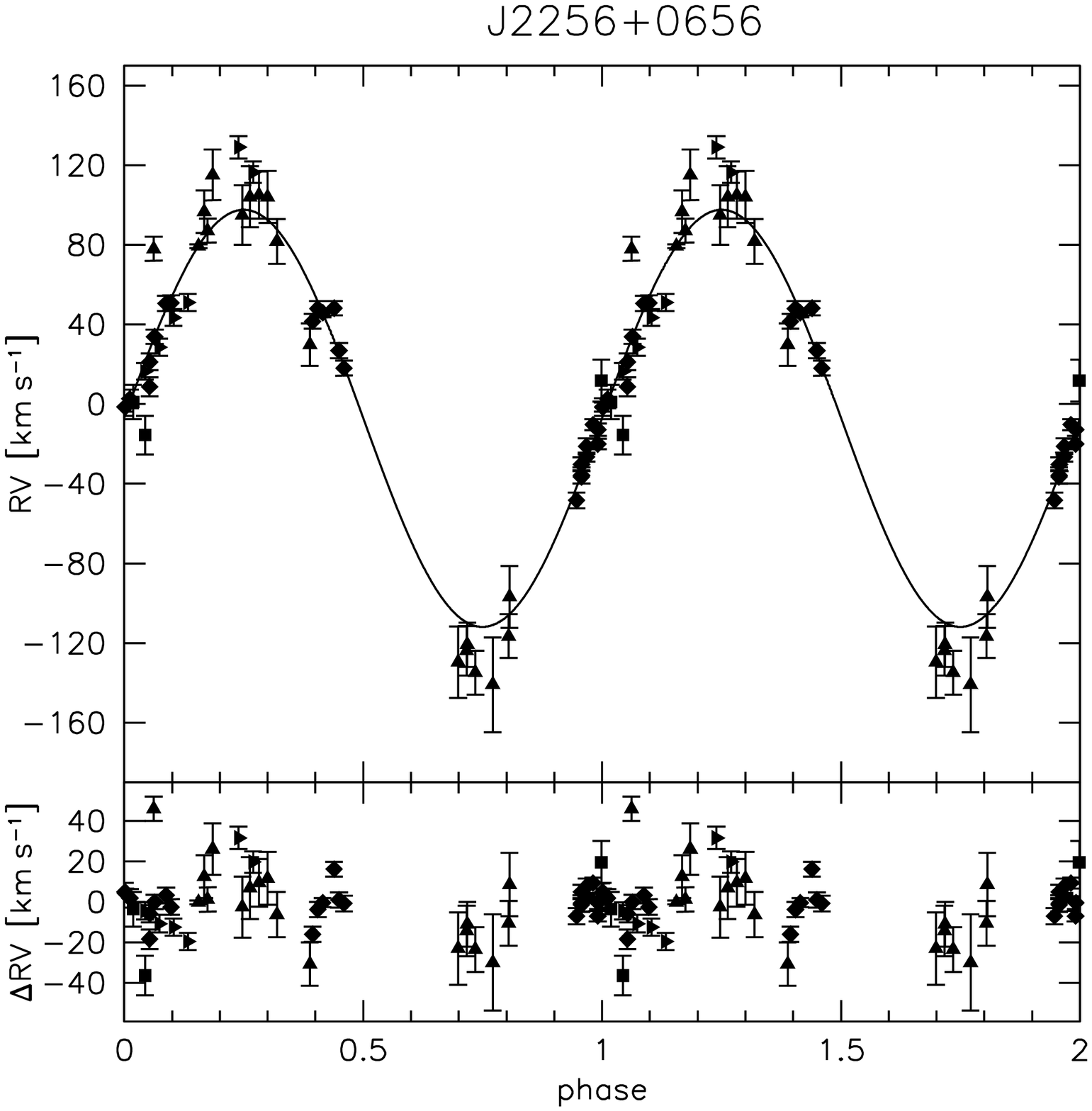}}
\end{center}
\caption{Radial velocity curves (see Fig~\ref{rv1}).}
\label{rv2}
\end{figure*}

\begin{table}[t!]
\caption{Follow-up observations 2009/2010.
        The first column lists the date of observation, while in the second
        the used telescope and instrumentation is shown. In the third column the observers are listed.}
\label{follow-up-runs}
\begin{center}
\begin{tabular}{llll} \hline
\noalign{\smallskip}
Date & Telescope\,\&\,Instrument & Observer\\ \hline
\noalign{\smallskip}
2009/06/05--2009/06/09 & ING-INT/IDS & R. \O., R. O., \\
& & T. O. \\
2009/07/22--2009/07/26 & CAHA-3.5m/TWIN & T. K. \\
2009/08/24--2009/08/27 & ING-WHT/ISIS & S. G. \\ 
2009/11/08--2009/11/12 & ESO-NTT/EFOSC2 & T. K. \\ 
April/August 2009 & Gemini-North/GMOS & Service \\ 
2010/02/12--2010/02/15 & SOAR/Goodman & B. B. \\ \hline
\end{tabular}
\end{center}
\end{table}

\section{Multi-site observations and data reduction}\label{s:data}
 
Follow-up medium resolution spectra were taken during de\-dicated follow-up runs (see Table~\ref{follow-up-runs}) with the EFOSC2 spectrograph ($R\simeq2200,\lambda=4450-5110\,{\rm \AA}$) mounted at the ESO\,NTT, the ISIS spectrograph ($R\simeq4000,\lambda=3440-5270\,{\rm \AA}$) mounted at the WHT, the TWIN spectrograph mounted at the CAHA-3.5m telescope ($R\simeq4000, \lambda=3460-5630\,{\rm \AA}$), the Goodman spectrograph mounted at the SOAR telescope ($R\simeq2500, \lambda=3500-6160\,{\rm \AA}$), the GMOS spectrograph ($R\simeq1200,\lambda=3770-4240\,{\rm \AA}$) mounted at the Gemini North telescope and the IDS spectrograph mounted at the Isaac Newton Telescope ($R\simeq1400,\lambda=3000-6800\,{\rm \AA}$). Informations about data  taken in the course of our survey are provided in Geier et al. (\cite{geier10c}). Additional data could be gathered, when our targets were observed with the IDS spectrograph (March 2007, observer: T. M., C. C.; $R\simeq4000, \lambda=3930-5100\,{\rm \AA}$) and the grating spectrograph (March 2003, April 2004, observer: T. M.; $R\simeq4600, \lambda=4170-5030\,{\rm \AA}$) mounted at the 1.9m Radcliffe Telescope. Example spectra are shown in Fig.~\ref{specexample}.

\begin{table*}[t!]
\caption{Derived orbital parameters.} 
\label{tab:orbits}
\begin{center}
\begin{tabular}{lllrr} 
Object & $T_{0}$ & P &  $\gamma$ & K\\
 & [$-$2\,450\,000] & [d] & [${\rm km\,s^{-1}}$] & [${\rm km\,s^{-1}}$]\\ 
\hline
\\[-3mm]
J0023$-$0029 & $5069.850\pm0.008$ & $1.4876\pm0.0001$ & $16.4\pm2.1$ & $81.8\pm2.9$ \\
J1138$-$0035 & $4991.388\pm0.001$ & $0.207536\pm0.000002$ & $23.3\pm3.7$ & $162.0\pm3.8$ \\
J1505+1108 & $4938.867\pm0.002$ & $0.74773\pm0.00005$ & $-77.1\pm1.2$ & $97.2\pm1.8$ \\
J1654+3037 & $4991.5322\pm0.0008$ & $0.25357\pm0.00001$ & $40.5\pm2.2$ & $126.1\pm2.6$ \\
J1726+2744 & $4981.667\pm0.005$ & $0.50198\pm0.00005$ & $-36.7\pm4.8$ & $118.9\pm3.7$ \\
J2046$-$0454 & $4693.352\pm0.002$ & $0.24311\pm0.00001$ & $87.6\pm5.7$ & $134.3\pm7.8$ \\
J2256+0656 & $5070.662\pm0.002$ & $0.7004\pm0.0001$ & $-7.3\pm2.1$ & $105.3\pm3.4$ \\
\hline \\[-3mm]
\end{tabular}
\end{center}
\end{table*}

In order to obtain a good wavelength calibration, arc lamp exposures have been taken before or after the single exposures. In addition to that bright single sdBs have been taken as RV standards in most of the runs. In some cases the RVs of certain instruments (TWIN, GMOS) had to be corrected by a constant offset of up to $\simeq50\,{\rm km\,s^{-1}}$, which was derived from the RV measurements of the standard stars. The slit width was always chosen to be smaller than the size of the seeing discs to minimize systematic errors due to movement of the objects within the slit. Reduction was done either with the \texttt{MIDAS}, \texttt{IRAF} or \texttt{PAMELA}\footnote{http://www2.warwick.ac.uk/fac/sci/physics/research/astro/people\\/marsh/software} and \texttt{MOLLY}$^{2}$ packages. 

\begin{table*}[t!]
\caption{Significance of the circular orbital solutions. The best solutions for the orbital periods are given together with their minimum $\chi^{2}$ and reduced $\chi^{2}$ values as well as the number $n$ of RVs. The second best aliases (further than $1\%$ away from the best solution) and the $\Delta \chi^{2}$-values with respect to the best solutions are given as well. The systematic error adopted to normalise the reduced $\chi^{2}$ ($e_{\rm norm}$) is given for each case.The probabilities for the orbital period to deviate from our best solution by more than $1\%$ ($p_{\rm false}[1\%]$) or $10\%$ ($p_{\rm false}[10\%]$) are given in the last columns.} 
\label{tab:sig}
\begin{center}
\begin{tabular}{llllllllll} 
Object & best solution & $\chi^{2}$ & $\chi^{2}_{\rm reduced}$ & 2nd best alias &  $\Delta\,\chi^{2}$ & $n$ & $e_{\rm norm}$ & $\log{p_{\rm false}}[1\%]$ & $\log{p_{\rm false}}[10\%]$ \\
       & [d] &  &  & [d] & & & [${\rm km\,s^{-1}}$] & & \\ 
\hline
\\[-3mm]
J0023$-$0029 & $1.4876$ & $157$ & $3.74$ & $0.5976$ & $130$ & $47$ & $8.0$ & $-3.0$    & $-3.4$    \\
J1138$-$0035 & $0.207536$ & $213$ & $5.33$ & $0.260192$ & $426$ & $45$ & $16.0$ & $-3.5$   & $-3.5$ \\
J1505+1108   & $0.74773$ & $155$ & $4.30$ & $0.75709$ & $679$ & $41$ & $7.0$ & $<-4.0$ &  $<-4.0$ \\
J1654+3037   & $0.25357$ & $18$  & $0.54$ & $0.20397$ & $64$ & $38$ & $-$  & $<-4.0$   &  \\
J1726+2744   & $0.50198$ & $82$  & $2.48$  & $1.00998$ & $77$ &  $38$  & $12.0$ & $-1.2$  & $-1.9$  \\
J2046$-$0454 & $0.24311$ & $52$  & $3.05$  & $0.31971$ & $39$ & $22$  & $17.0$ & $-1.1$   & $-1.1$   \\
J2256+0656   & $0.7004$ & $276$ & $6.13$  & $2.1903$  & $976$ &  $50$ & $13.0$ & $<-4.0$  & $<-4.0$  \\
\hline \\[-3mm]
\end{tabular}
\end{center}
\end{table*}

\section{Orbital parameters \label{s:orbit}}

\begin{figure*}[t!]
\begin{center}
        \resizebox{6.0cm}{!}{\includegraphics{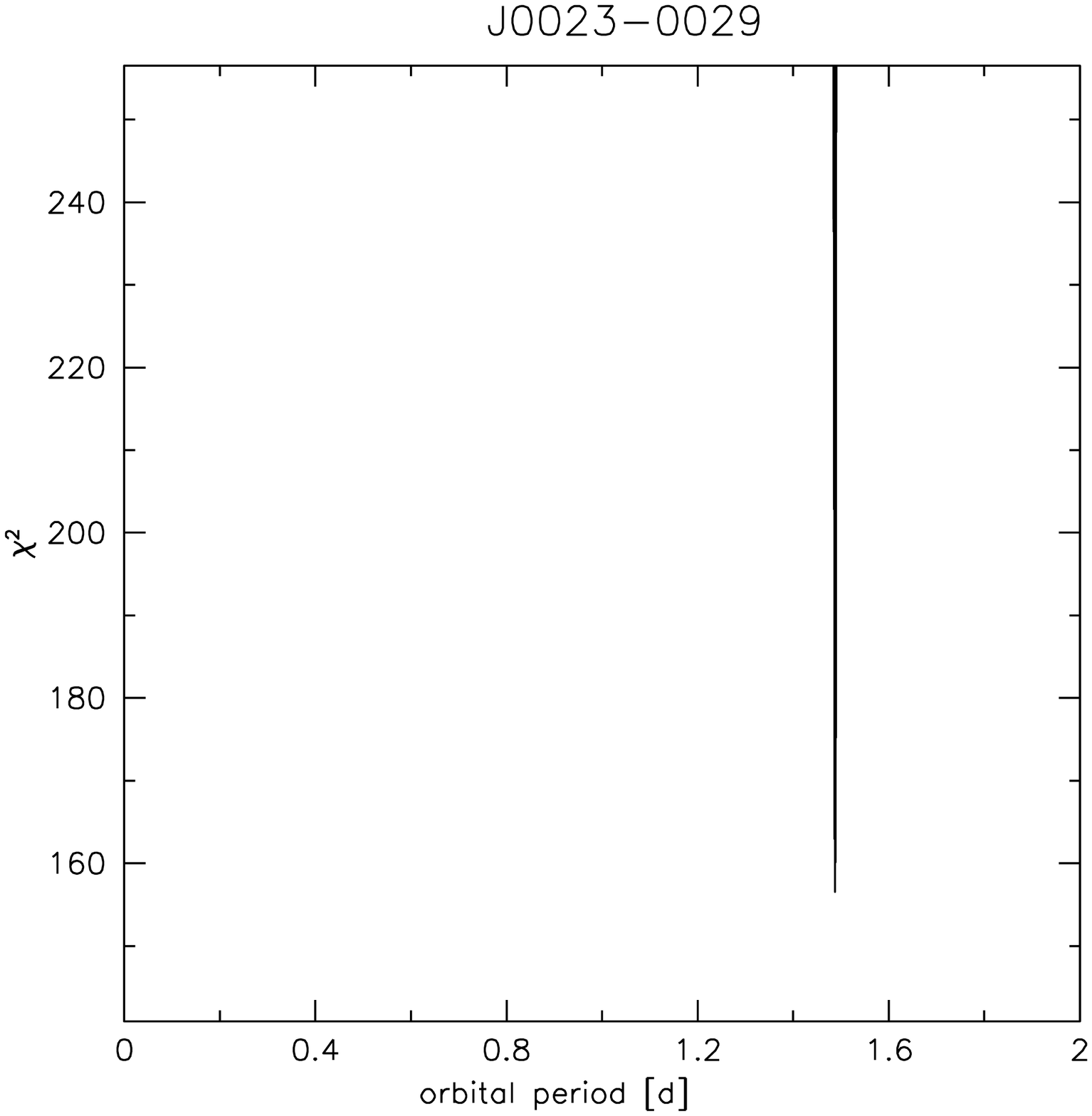}}
        \resizebox{6.0cm}{!}{\includegraphics{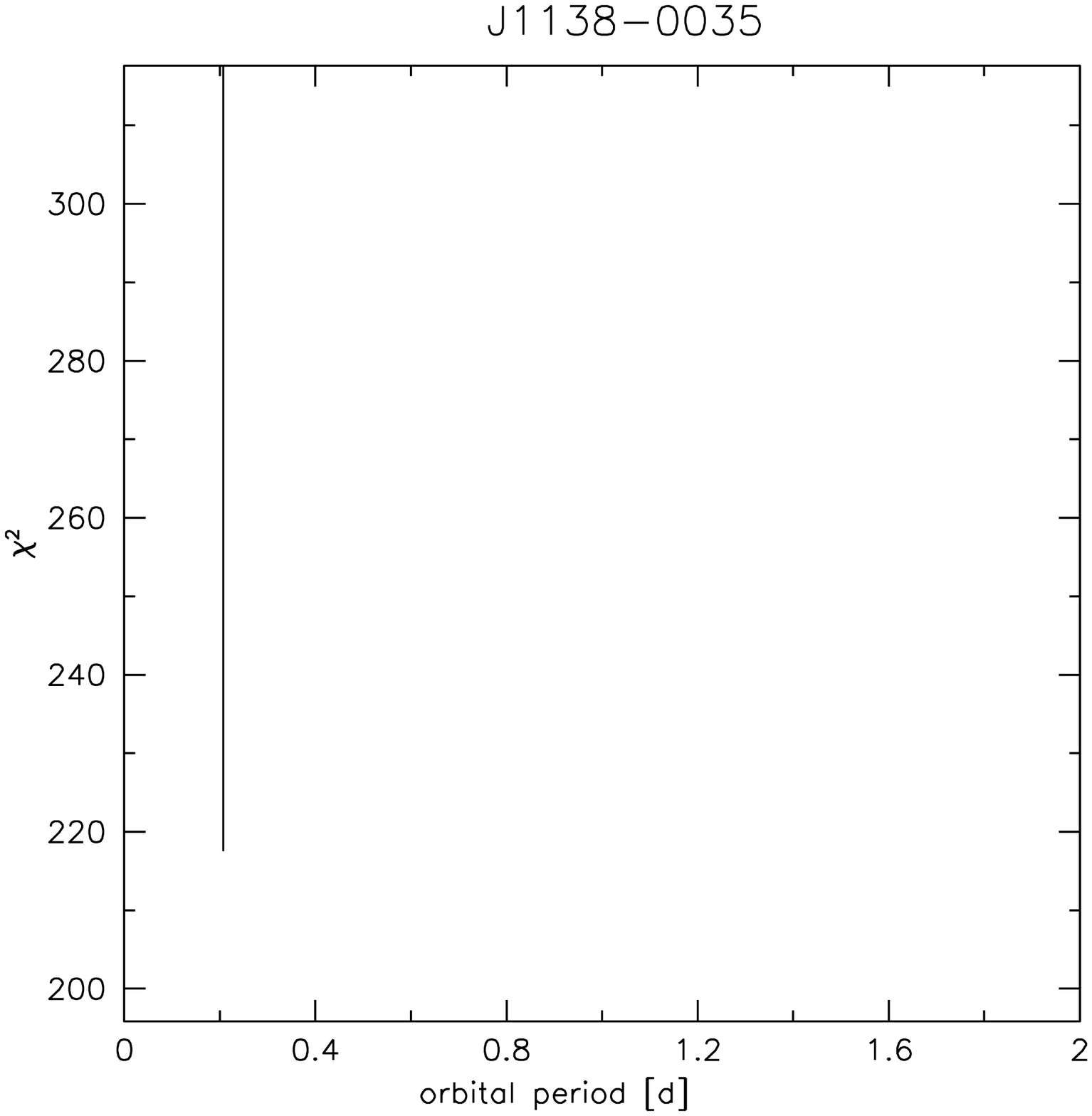}}
	\resizebox{6.0cm}{!}{\includegraphics{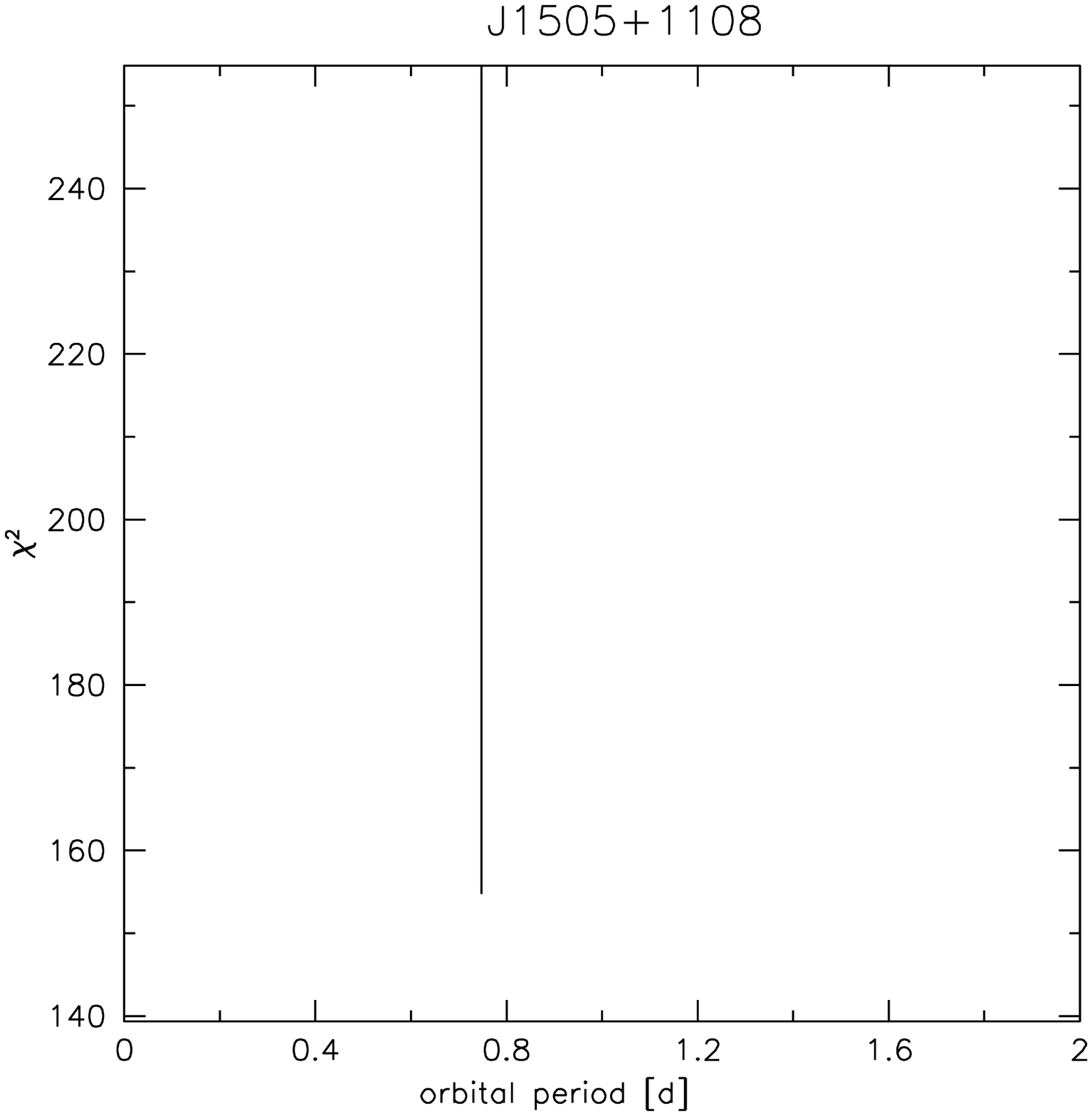}}
	\resizebox{6.0cm}{!}{\includegraphics{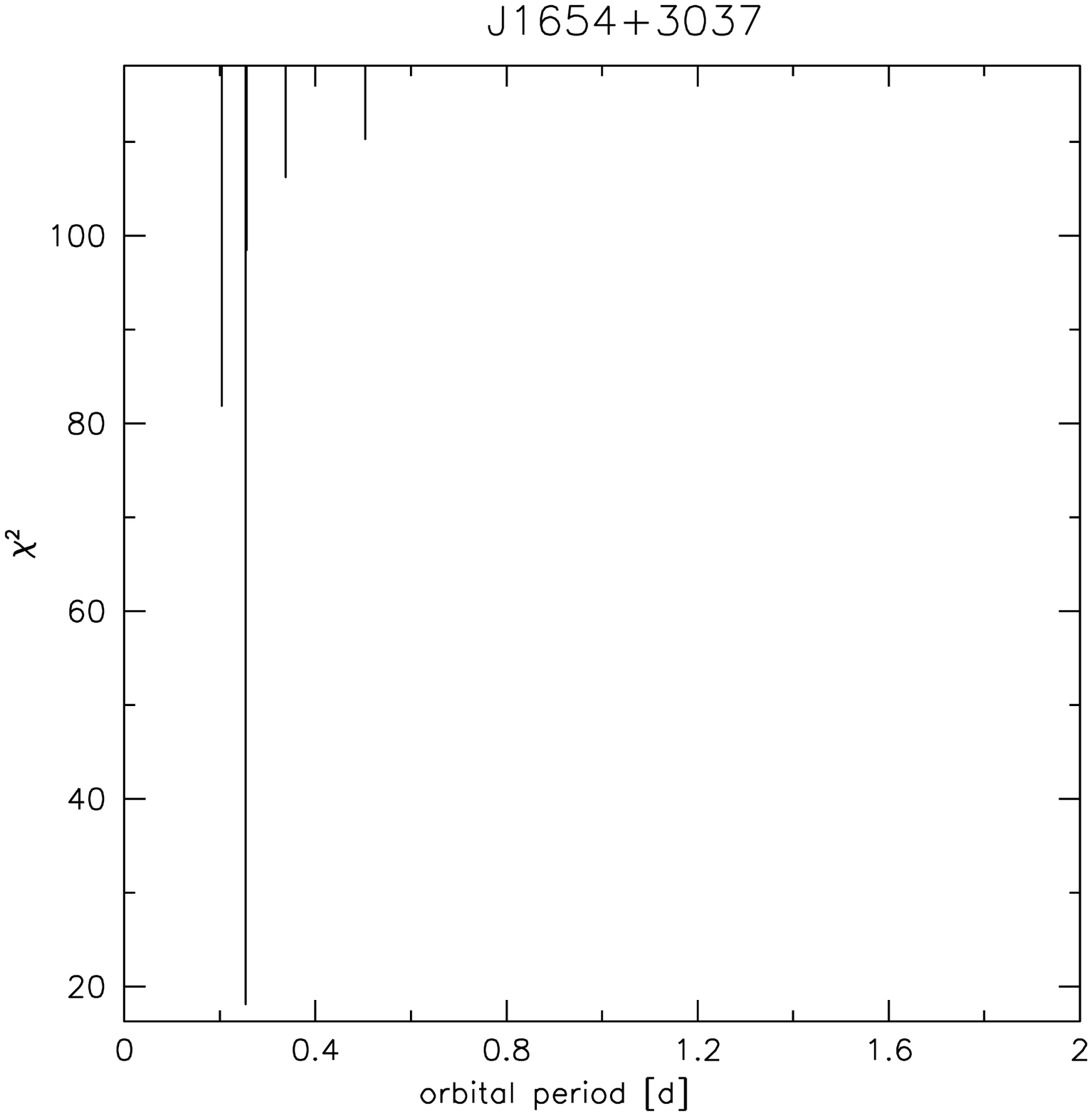}}
	\resizebox{6.0cm}{!}{\includegraphics{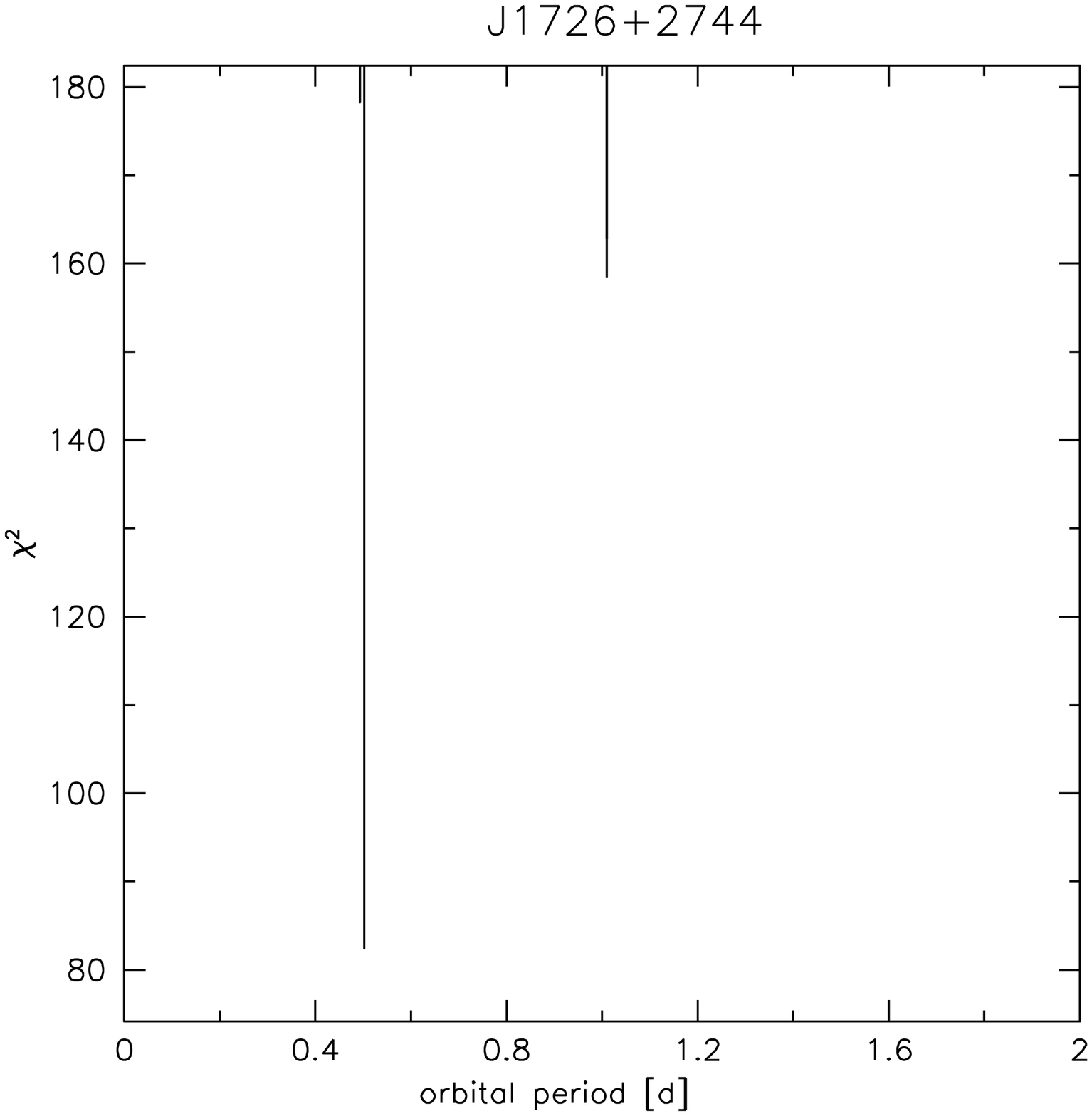}}
        \resizebox{6.0cm}{!}{\includegraphics{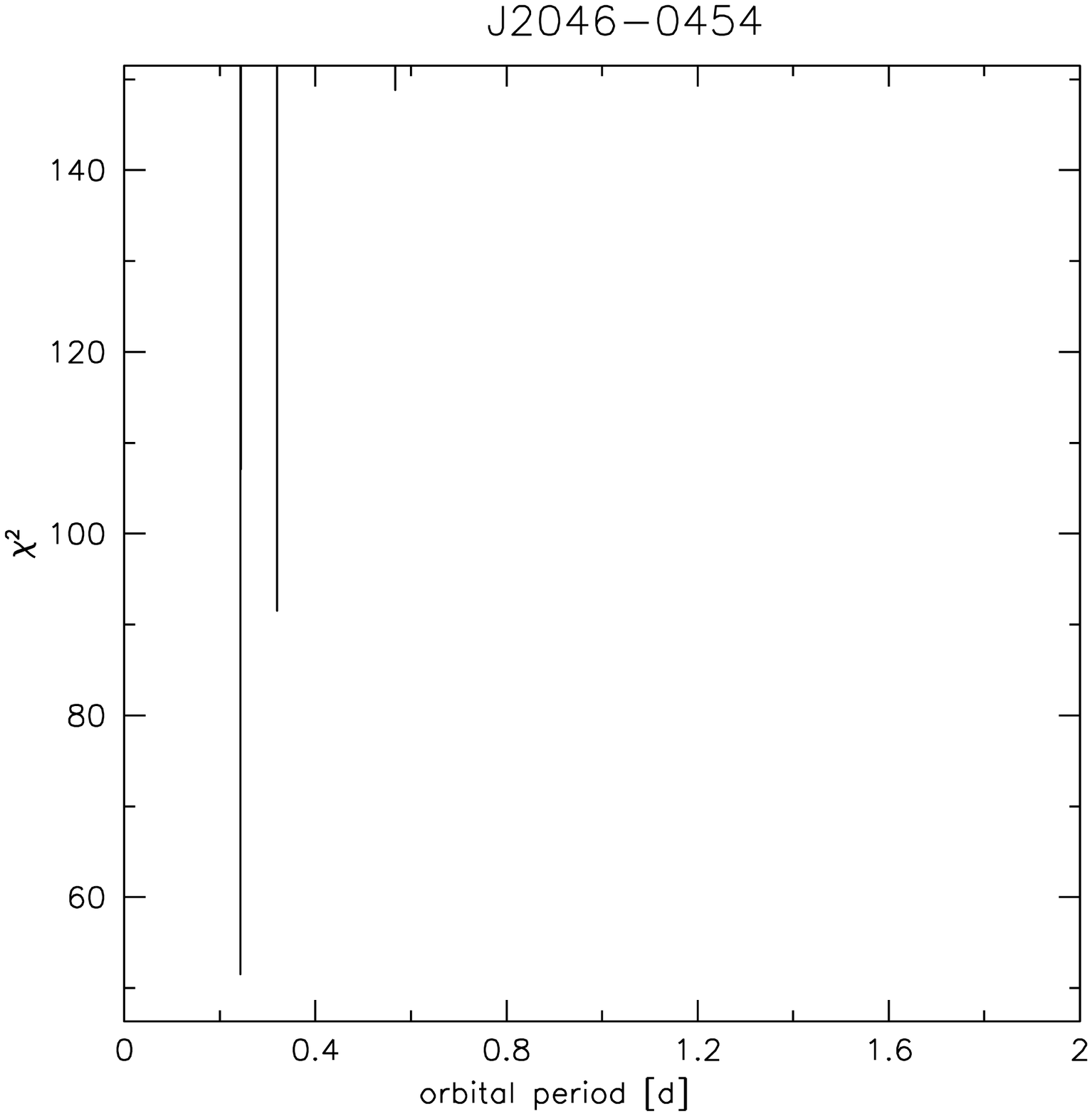}}
        \resizebox{6.0cm}{!}{\includegraphics{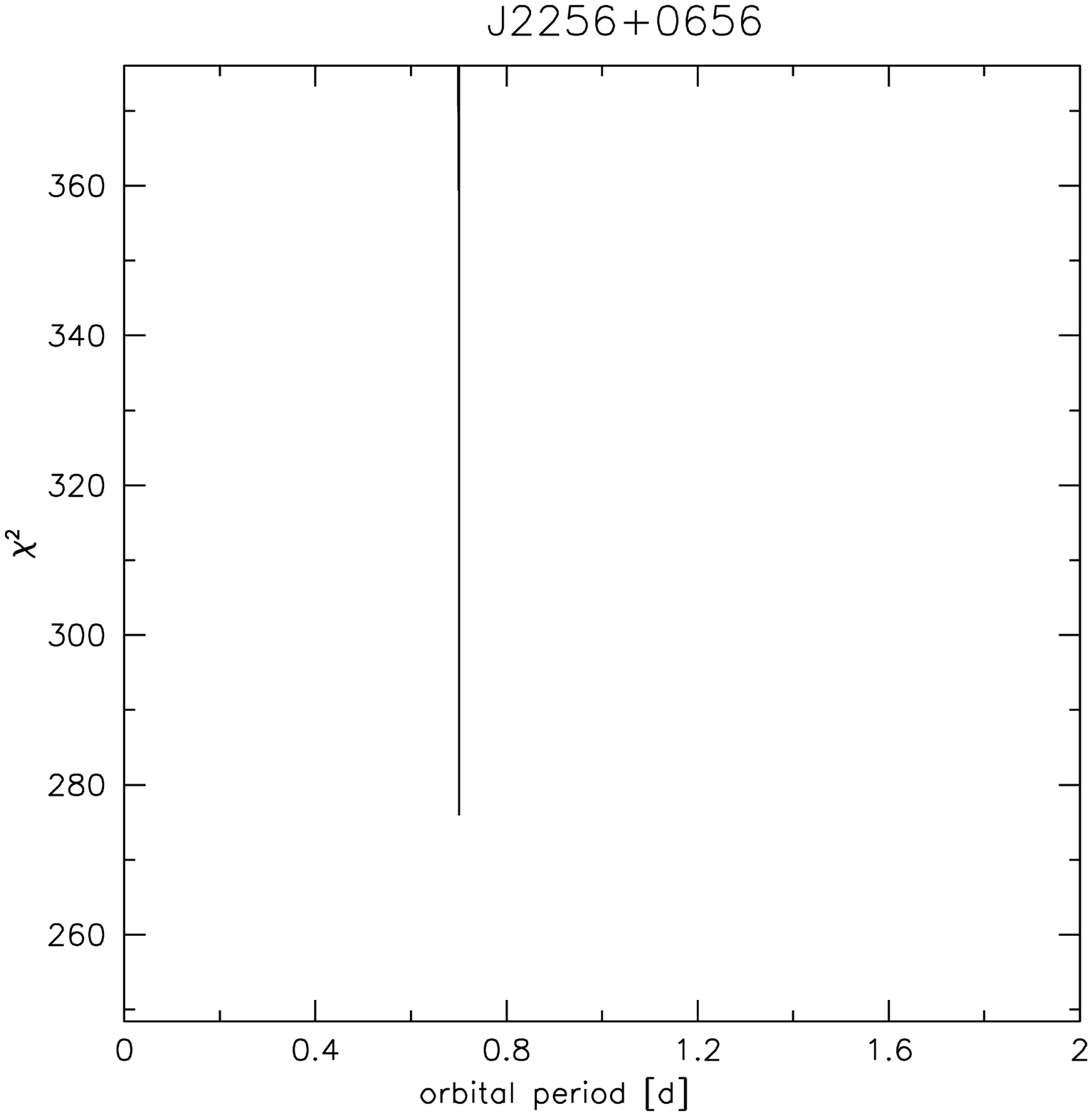}}
\end{center}
\caption{$\chi^{2}$ plotted against orbital period. The lowest peak corresponds to the most likely solution.}
\label{chi}
\end{figure*}

The radial velocities were measured by fitting a set of mathematical functions (Gaussians, Lorentzians and polynomials) to the hydrogen Balmer lines as well as helium lines if present using the FITSB2 routine (Napiwotzki et al. \cite{napiwotzki04b}). The RVs of the GMOS spectra have been measured by fitting three Gaussians to the $H_{\rm \gamma}$ line. Three functions are used to match the continuum, the line and the line core, respectively and mimic the typical Voigt profile of spectral lines. The profiles are fitted to all suitable lines simultaneously using $\chi^{2}$-minimization and the RV shift with respect to the rest wavelengths is measured. The RVs and formal $1\sigma$-errors are given in Appendix~\ref{app:RV}. Assuming circular orbits sine curves were fitted to the RV data points in fine steps over a range of test periods. For each period the $\chi^{2}$ of the best fitting sine curve was determined. The result is similar to a power spectrum with the lowest $\chi^{2}$ indicating the most likely period (see Fig.~\ref{chi}). 
In order to estimate the significance of the orbital solutions and the contributions of systematic effects to the error budget, we normalised the $\chi^{2}$ of the most probable solution by adding systematic errors in quadrature until the reduced $\chi^{2}$ reached $\simeq1.0$. Using these modified uncertainties we performed Monte Carlo simulations for the most likely periods. For each simulation a randomised set of RVs was drawn from Gaussian distributions with central value and width corresponding to the RV measurements and the analysis repeated. From these simulations the probabilities for the orbital periods to deviate from our best solution by more than $1\%$ or $10\%$ were calculated.

In order to derive most conservative errors for the RV semi-amplitude $K$ and the system velocity $\gamma$ we fixed the most likely period and created new RV datasets with a bootstrapping algorithm. Ten thousand RV datasets were obtained by random sampling with replacement from the original dataset. In each case an orbital solution was calculated in the way described above. The standard deviation of these results was adopted as error estimate. The RV curves are given in Figs.~\ref{rv1} and  \ref{rv2}. The residuals of the RV curves after subtracting the best orbital solution are of the same order in all cases (see Figs.~\ref{rv1}, \ref{rv2}). The accuracy is limited by the resolution of the spectra and their signal-to-noise. Combining data obtained with different instruments is also expected to contribute to the systematic error. Nevertheless, we found that all orbital solutions given here are significant  (see Table~\ref{tab:sig}).

Edelmann et al. (\cite{edelmann05}) reported the discovery of small eccentricities ($e<0.06$) in the orbital solutions of five close hot subdwarf binaries. All of these binaries are expected to have formed via common envelope ejection. Although the CE phase is very short, it should nevertheless be very efficient in circularising the binary orbits. That is why the discovery of Edelmann et al. (\cite{edelmann05}) came as a surprise. Napiwotzki et al. (in prep.) found more such systems with even shorter periods.

In order to investigate whether the orbital solutions of our programme binaries can be improved by allowing for eccentricity, we fitted eccentric orbits to our radial velocity data and performed statistical tests (F-test, see Pringle \cite{pringle75}, and the Bayesian information criterion BIC) to check whether eccentric solutions are significant or not. In all cases the circular solutions were preferred. However, the derived upper limits for the orbital eccentricities range from $0.15$ to $0.3$, which means that low eccentricities as the ones reported by Edelmann et al. (\cite{edelmann05}) cannot be firmly excluded.

\section{Atmospheric parameters \label{s:atmo}}

Atmospheric parameters have been determined by fitting model spectra to the hydrogen Balmer and helium lines in the way described in Geier et al. (\cite{geier07}). The single spectra have been corrected for their orbital motion and coadded. Depending on the effective temperature of the stars, LTE models with solar metallicity ($T_{\rm eff}<30\,000\,{\rm K}$) or ten times solar metallicity ($T_{\rm eff}>30\,000\,{\rm K}$) have been used. The enhanced metallicity models account for the radiative levitation of heavy elements in the diffusion dominated atmospheres (for a detailed discussion see O'Toole \& Heber \cite{otoole06}). 

In order to investigate systematic effects introduced by the individual instruments, especially the different resolutions and wavelength coverages, the parameters have been derived separately from spectra taken with different instruments. As can be seen in Table~\ref{tab:atm} no constant systematic shifts are present. The weighted means have been calculated and adopted as final solutions. Typical systematic errors introduced by different model grids are of the order of $\pm0.05$ in $\log{g}$ and $500\,{\rm K}$ in $T_{\rm eff}$ (e.g. Lisker et al. \cite{lisker05}; Geier et al. \cite{geier07}). These uncertainties were added in quadrature to the statistical errors.

Three of our programme stars have been classified as hot subdwarfs by Eisenstein et al. (\cite{eisenstein06}), but the authors pointed out that the atmospheric parameters of the sdO/Bs given in their catalogue are not accurate.

\begin{figure}[t!]
\begin{center}
	\resizebox{8.5cm}{!}{\includegraphics{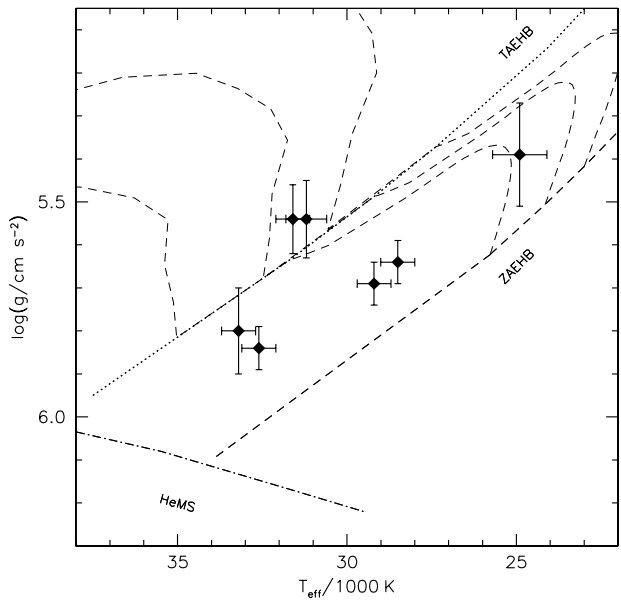}}
\end{center}
\caption{$T_{\rm eff}-\log{g}$-diagram. The helium main sequence (HeMS) and the EHB band (limited by the zero-age EHB, ZAEHB, and the terminal-age EHB, TAEHB) are superimposed with EHB evolutionary tracks from Dorman et al. (\cite{dorman93}).}
\label{tefflogg}
\end{figure}

All stars of our sample are situated on or near the Extreme Horizontal Branch (EHB) and are most likely core-helium burning stars (see Fig.~\ref{tefflogg}). Since the orbital periods of these binaries are short, they can only have formed via common envelope ejection. Population synthesis models (Han et al. \cite{han02}, \cite{han03}) predict a mass range of $M_{\rm sdB}=0.37-0.48\,M_{\rm \odot}$ for sdBs in binaries formed in this way. The mass distribution shows a sharp peak at a mass of about $0.47\,{\rm M_{\odot}}$. This theoretical mass distribution is consistent with analyses of close binary systems (e.g. Geier et al. \cite{geier07}; For et al. \cite{for10}) as well as asteroseismic analyses of pulsating sdBs (see Charpinet et al. \cite{charpinet08} and references therein). If the progenitor star was massive enough on the main sequence to ignite core helium-burning under non-degenerate conditions, the sdB mass may be as low as $0.3\,{\rm M_{\odot}}$. A small fraction of the sdB population is predicted to be formed in that way (Han et al. \cite{han02}, \cite{han03}). Especially for sdB binaries with massive companions this formation scenario may become important.

\section{Constraining the nature of the unseen companions \label{s:comp}}

Since the programme stars are single-lined spectroscopic binaries, only their mass functions can be calculated.

 \begin{equation}
 \label{equation-mass-function}
 f_{\rm m} = \frac{M_{\rm comp}^3 \sin^3i}{(M_{\rm comp} +
   M_{\rm sdB})^2} = \frac{P K^3}{2 \pi G}
 \end{equation}

Although the RV semi-amplitude $K$ and the period $P$ can be derived from the RV curve, the sdB mass $M_{\rm sdB}$, the companion mass $M_{\rm comp}$ and the inclination angle $i$ remain free parameters. Adopting $M_{\rm sdB}=0.47\,{\rm M_{\odot}}$ and $i<90^{\rm \circ}$ we derive a lower limit for the companion mass (see Table\,\ref{rvmasses}).

For mini\-mum companion masses lower than $0.45\,M_{\rm \odot}$ the companion may be a late type main sequence star or a compact object like a WD. Main sequence stars in this mass range are outshined by the sdBs and not visible in optical spectra (Lisker et al. \cite{lisker05}). That is the reason why the companions' nature still remains unknown for most of the $\simeq$80 known sdB systems with low minimum companion masses (see Fig.~\ref{periodK}). If on the other hand the minimum companion mass exceeds $0.45\,M_{\rm \odot}$, spectral features of a main sequence companion become visible in the optical. The non-detection of such features therefore allows us to exclude a main sequence star. The companion must then be a compact object. More massive compact companions like massive WDs, neutron stars or black holes are more likely as soon as the minimum mass exceeds $1.00\,M_{\rm \odot}$ or even the Chandrasekhar limit $1.40\,M_{\rm \odot}$.

Due to the fact that we selected targets with high RV shifts, the distribution of orbital inclinations in our target sample is not random any more. Our selection strategy strongly favours high inclination angles, and therefore the companion masses are likely to be close to their minimum values. The probability of detecting eclipses, reflection effects or variations caused by ellipsoidal deformation in the light curves of systems with short orbital periods should therefore be significantly higher than in an unbiased sample. 

\begin{table}[t!]
\caption{Derived minimum masses and most probable nature of the companions.} 
\label{rvmasses}
\begin{center}
\begin{tabular}{llll} 
Object & $f(M)$ & $M_{\rm 2min}$ & Companion\\
 & [$M_{\rm \odot}$] & [$M_{\rm \odot}$] &  \\ 
\hline
\\[-3mm]
J0023$-$0029 & $0.084$ & $0.40$ & MS/WD \\
J1138$-$0035 & $0.091$ & $0.42$ & WD \\
J1505+1108 & $0.071$ & $0.37$   & MS/WD \\
J1654+3037 & $0.053$ & $0.32$   & MS/WD \\
J1726+2744 & $0.087$ & $0.41$    & MS/WD \\
J2046$-$0454 & $0.061$ & $0.34$  & MS/WD \\
J2256+0656 & $0.085$ & $0.40$    & MS/WD \\
\hline \\[-3mm]
\end{tabular}
\end{center}
\end{table}

\section{Results}\label{s:results}

The spectra of all stars in our sample have been checked for spectral features of their companions. Hot subdwarfs with faint main sequence companions usually show spectral lines of the Mg\,{\sc i} triplet at $\simeq5170\,{\rm \AA}$ (Lisker et al. \cite{lisker05}) and the Ca\,{\sc ii} triplet at $\simeq8650\,{\rm \AA}$. No such features are visible in the spectra of our programme stars (see e.g. Fig.~\ref{specexample}). Stark \& Wade (\cite{stark03}) ana\-lysed optical and IR photometry (2MASS) and found no indication of an IR-excess caused by a cool companion in the case of J1654+3037. According to the catalogue of Reed \& Stiening (\cite{reed04}), who performed a similar analysis, J1505+1108 shows signs of an IR-excess in the H and K-bands, but the large errors of these measurements and the missing spectral signatures of a cool companion in the SDSS spectra are strong indications, that no visible companion is present. 

J1654+3037 and J2046$-$0454 have very similar orbital parameters. The periods are short ($0.25\,{\rm d}$) and the minimum companion masses are constrained to $0.32\,{\rm M_{\odot}}$ and $0.34\,{\rm M_{\odot}}$. Whether the companions are M dwarfs or WDs is therefore not yet clear. In the former case a reflection effect should be easily detectable in the light curves. Photometric follow-up will allow us to clarify the nature of the companions.

\begin{figure}[t!]
\resizebox{8.5cm}{!}{\includegraphics{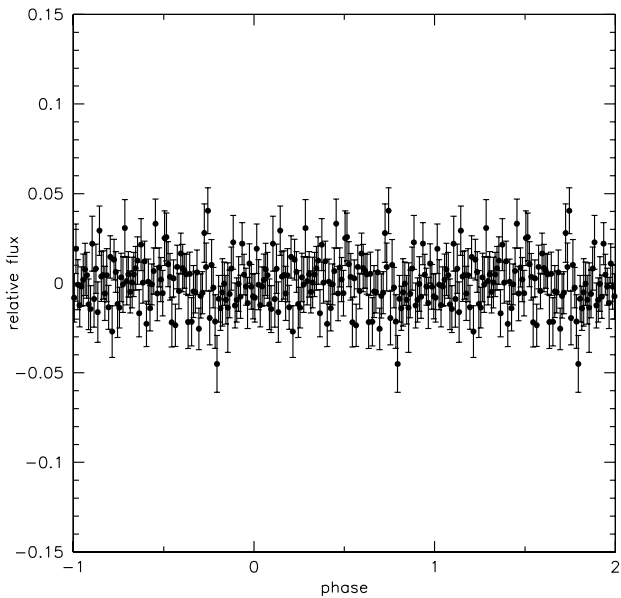}}
\caption{SuperWASP light curve of J1138$-$0035 folded to the orbital phase. The 11213 data points taken between 2006/07/05 and 2009/07/02 are binned to 100 phase bins. Relative flux is plotted against the orbital phase.}
\label{lc}
\end{figure}

The companion of the short period ($0.2\,{\rm d}$) system J1138$-$0035 is most likely a white dwarf. The minimum companion mass is constrained to $0.42\,{\rm M_{\odot}}$ and no sign of a companion is seen in the spectra. A light curve taken by the SuperWASP project (Pollacco et al. \cite{pollacco06}) shows no variation exceeding $\simeq1\%$ (see Fig.~\ref{lc}). Due to the short period of this system a reflection effect should be visible, if the companion should be a cool main sequence star. The absence of such a variation leads to the conclusion that the companion is most likely a white dwarf.

The orbital periods of J1726+2744 ($0.5\,{\rm d}$), J2256+0656 ($0.7\,{\rm d}$) and J1505+1108 ($0.74\,{\rm d}$) are longer. Their minimum companion masses are similar ($0.37-0.41\,{\rm M_{\odot}}$) and close to the border between main sequence stars and white dwarfs. The companions of J1726+2744 and J2256+0656 are most likely WDs. Koen (\cite{koen09}) and Shimanskii et al. (\cite{shimanskii08}) recently showed that reflection effects can still be detected in the light curves of sdB binaries with similar orbital periods. A reflection effect in J0023$-$0029 on the other hand is most likely not detectable, because the orbital period is too long ($1.5\,{\rm d}$). 

\section{Efficiency of target selection}\label{s:efficient}

\begin{figure}[t!]
	\resizebox{\hsize}{!}{\includegraphics{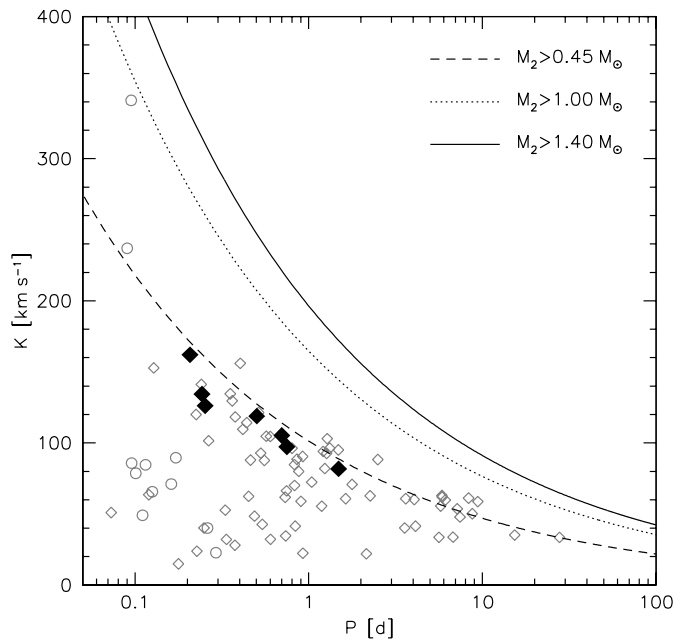}}
	\caption{The RV semiamplitudes of all known sdB binaries with spectroscopic solutions plotted against their orbital periods (Geier et al. \cite{geier10c}). Binaries which have initially been discovered in photometric surveys due to indicative features in their light curves (eclipses, reflection effects, ellipsoidal variations) are marked with open circles. Binaries discovered by detection of RV variations from time resolved spectroscopy are marked with open diamonds. The dashed, dotted and solid lines mark the regions to the right where the minimum companion masses derived from the binary mass function (assuming $0.47\,{\rm M_{\odot}}$ for the sdBs) exceed $0.45\,{\rm M_{\odot}}$, $1.00\,{\rm M_{\odot}}$ and $1.40\,{\rm M_{\odot}}$. The seven binaries from the MUCHFUSS project are marked with filled diamonds.}
\label{periodK}
\end{figure}

The goal of the MUCHFUSS project is to find sdB binaries with massive compact companions and study this population of close binaries. We tried to optimise our target selection to achieve this goal. Fig.~\ref{periodK} illustrates the efficiency of our target selection. The RV semiamplitudes of all known sdB binaries with spectroscopic solutions (open symbols) are plotted against their orbital periods (Geier et al. \cite{geier10c}). Binaries which have initially been discovered in photometric surveys due to indicative features in their light curves (eclipses, reflection effects, ellipsoidal variations) are marked with open circles. Binaries discovered by RV variations from time resolved spectroscopy are marked with open diamonds. The dashed, dotted and solid lines mark the regions to the right where the minimum companion masses derived from the binary mass function (assuming $0.47\,{\rm M_{\odot}}$ for the sdBs) exceed $0.45\,{\rm M_{\odot}}$, $1.00\,{\rm M_{\odot}}$ and $1.40\,{\rm M_{\odot}}$.

Most of the known sdB binaries are situated beneath the $0.45\,{\rm M_{\odot}}$ line, which means that the companion type cannot be constrained from the mass function alone. Photometry is necessary to clarify the companions' nature in these cases. The most massive sdB binary known to date is KPD\,1930+2752 with a WD  companion of $0.9\,{\rm M_{\odot}}$. This short period system has been discovered based on indicative features in its light curve (upper left corner in Fig.~\ref{periodK}; Bill\`{e}res et al. \cite{billeres00}).

The seven binaries from the MUCHFUSS project are marked with filled diamonds. It can be clearly seen that they belong to the sdB binary population with the largest minimum masses close to $0.45\,{\rm M_{\odot}}$. We therefore conclude that our target selection is efficient and singles out sdB binaries with massive companions.

\section{Summary and Outlook}\label{s:summary}

A multi-site follow-up campaign is being conducted with medium resolution spectrographs mounted at several different telescopes of mostly $2\,{\rm m}$ to $4\,{\rm m}$-class. First results were presented for seven close binary sdBs with short orbital periods ranging from $\simeq0.21\,{\rm d}$ to $1.5\,{\rm d}$ and most likely compact companions. The atmospheric parameters of all objects are compatible with core helium-burning stars on the EHB. Comparing our small sample with the known population of close sdB binaries we are able to show that our target selection method is efficient. All binaries solved up to now have high minimum companion masses compared to the rest of the sdB binary population. 

Up to now we have found significant orbital solutions for about $10\%$ of our target sample. Photometric follow-up observations will allow us to clarify the nature of the companions in most cases. A database of more than $700$ spectra has been built up and some binaries will be solvable with only a few additional RV points. 

\begin{acknowledgements}

A.T., S.G. and H.H. are supported by the Deutsche Forschungsgemeinschaft (DFG) through grants HE1356/45-1, HE1356/49-1, and HE1356/44-1, 
respectively. R.\O. acknowledges funding from the European Research Council under the European Community's Seventh Framework Programme (FP7/2007--2013)/ERC grant agreement N$^{\underline{\mathrm o}}$\,227224 ({\sc prosperity}), as well as from the Research Council of K.U.Leuven grant agreement GOA/2008/04. Travel to the DSAZ (Calar Alto, Spain) was supported by DFG under grants HE1356/48-1 and HE1356/50-1. Travel to La Palma for the observing run at the WHT was funded by DFG through grant He 1356/53-1.

Funding for the SDSS and SDSS-II has been provided by the Alfred P. Sloan Foundation, the Participating Institutions, the National Science Foundation, the U.S. Department of Energy, the National Aeronautics and Space Administration, the Japanese Monbukagakusho, the Max Planck Society, and the Higher Education Funding Council for England. The SDSS Web Site is http://www.sdss.org/.

The SDSS is managed by the Astrophysical Research Consortium for the Participating Institutions. The Participating Institutions are the American Museum of Natural History, Astrophysical Institute Potsdam, University of Basel, University of Cambridge, Case Western Reserve University, University of Chicago, Drexel University, Fermilab, the Institute for Advanced Study, the Japan Participation Group, Johns Hopkins University, the Joint Institute for Nuclear Astrophysics, the Kavli Institute for Particle Astrophysics and Cosmology, the Korean Scientist Group, the Chinese Academy of Sciences (LAMOST), Los Alamos National Laboratory, the Max-Planck-Institute for Astronomy (MPIA), the Max-Planck-Institute for Astrophysics (MPA), New Mexico State University, Ohio State University, University of Pittsburgh, University of Portsmouth, Princeton University, the United States Naval Observatory, and the University of Washington. 

\end{acknowledgements}

\newpage

\begin{appendix}

\section{Atmospheric parameters}

\begin{table}[h!]
\caption{Atmospheric parameters} \label{tab:atm}
\begin{center}
\begin{tabular}{lllll}
\hline
\noalign{\smallskip}
Object & $T_{\rm eff}$ & $\log{g}$ & $\log{y}$ & Instrument \\
       & [K] &  &  &  \\ 
\noalign{\smallskip}
\hline
\noalign{\smallskip}
J0023$-$0029 & $30100\pm500$ & $5.62\pm0.08$ & $-2.0$ & SDSS \\
                     & $29000\pm100$ & $5.71\pm0.02$ & $-2.0$ & ISIS \\
                     & $29200\pm500$ & $5.69\pm0.05$ & $-2.0$ & adopted \\
\noalign{\smallskip}
\hline
\noalign{\smallskip}
J1138$-$0035 & $30800\pm500$ & $5.50\pm0.09$ & $-3.0$ & SDSS \\
                     & $31700\pm700$ & $5.59\pm0.10$ & $-3.0$ & IDS \\
                     & $31200\pm600$ & $5.54\pm0.09$ & $-3.0$ & adopted \\
\noalign{\smallskip}
\hline
\noalign{\smallskip}
J1505+1108 & $33300\pm500$ & $5.80\pm0.10$ & $-2.4$ & SDSS \\
                     & $33000\pm600$ & $5.80\pm0.11$ & $-2.2$ & TWIN \\
                     & $33200\pm500$ & $5.80\pm0.10$ & $-2.3$ & adopted \\
\noalign{\smallskip}
\hline
\noalign{\smallskip}
J1654+3037 & $24400\pm800$ & $5.32\pm0.11$ & $-2.3$ & SDSS \\
                     & $25500\pm900$ & $5.47\pm0.13$ & $-2.5$ & IDS \\  
                     & $24900\pm800$ & $5.39\pm0.12$ & $-2.4$ & adopted \\
\noalign{\smallskip}
\hline
\noalign{\smallskip}
J1726+2744 & $33500\pm400$ & $5.71\pm0.09$ & $-2.2$ & SDSS \\
                     & $33300\pm400$ & $5.91\pm0.06$ & $-2.2$ & TWIN \\
                     & $32300\pm100$ & $5.87\pm0.02$ & $-2.2$ & ISIS \\
                     & $32400\pm700$ & $5.73\pm0.12$ & $-2.1$ & IDS \\
                     & $32600\pm500$ & $5.84\pm0.05$ & $-2.2$ & adopted \\
\noalign{\smallskip}
\hline
\noalign{\smallskip}
J2046$-$0454 & $31600\pm600$ & $5.55\pm0.10$ & $-3.0$ & SDSS \\
                     & $32100\pm500$ & $5.57\pm0.09$ & $-3.0$ & TWIN \\
                     & $31100\pm400$ & $5.52\pm0.06$ & $-3.0$ & FORS1 \\
                     & $31600\pm500$ & $5.54\pm0.08$ & $-3.0$ & adopted \\
\noalign{\smallskip}
\hline
\noalign{\smallskip}
J2256+0656 & $28900\pm600$ & $5.58\pm0.11$ & $-3.0$ & SDSS \\
                     & $29200\pm900$ & $5.74\pm0.09$ & $-2.2$ & TWIN \\
                     & $28400\pm100$ & $5.63\pm0.02$ & $-2.2$ & ISIS \\
                     & $28500\pm500$ & $5.64\pm0.05$ & $-2.3$ & adopted \\
\noalign{\smallskip}
\hline
\end{tabular}
\end{center}
\end{table}

\vspace{10cm}
\section{Radial velocities}\label{app:RV}

\begin{table}[h!]
\caption{J0023$-$0029}
\label{RV1}
\begin{center}
\begin{tabular}{lrl}
\hline
\noalign{\smallskip}
mid$-$HJD & RV [${\rm km\,s^{-1}}$] & Instrument\\
$-2\,450\,000$ & & \\
\noalign{\smallskip}
\hline
\noalign{\smallskip}
1816.782390 & 92.0 $\pm$ 10.0 & SDSS \\
1816.794497 &     94.1 $\pm$    6.0 & \\
1816.806476 &     92.7 $\pm$     7.0 & \\
1885.575567 &      5.6 $\pm$     9.0 & \\
1885.587847 &      2.2 $\pm$    11.0 & \\
1885.600718 &      2.4 $\pm$    15.0 & \\
1885.614294 &     -3.1 $\pm$     9.0 & \\
1885.627153 &      7.7 $\pm$     9.0 & \\
1885.639444 &     -9.0 $\pm$     8.0 & \\
1899.578750 &    -36.7 $\pm$    14.0 & \\
1899.590972 &    -17.2 $\pm$     9.0 & \\
1899.603316 &    -27.7 $\pm$    10.0 & \\
1899.615509 &    -12.1 $\pm$    11.0 & \\
1900.573652 &    -13.8 $\pm$     7.0 & \\
1900.585712 &     -9.6 $\pm$    22.0 & \\
1900.599109 &    -25.3 $\pm$    16.0 & \\
\noalign{\smallskip}
\hline
\noalign{\smallskip}
5068.61892  &   90.6  $\pm$    3.0   & ISIS \\
5068.62614  &   94.8   $\pm$    3.7    & \\
5068.63335  &   81.4   $\pm$    4.0    & \\ 
5068.65013  &  102.0   $\pm$    4.0    & \\ 
5068.65736  &  103.5   $\pm$    4.4    & \\ 
5068.66457  &   94.9   $\pm$    3.7   & \\ 
5069.69141  &  -25.4   $\pm$    4.6   & \\  
5069.69862  &  -65.8   $\pm$    4.1   & \\  
5069.70585  &  -32.1   $\pm$    2.9   & \\ 
5069.71578  &  -28.2   $\pm$    2.6    & \\ 
5069.72300  &  -17.9   $\pm$   4.3    & \\
5069.73023  &   -6.2   $\pm$    4.1   & \\
5071.59347  &   71.9   $\pm$    5.9   & \\  
5071.60068  &   79.3   $\pm$    6.4   & \\ 
5071.60442  &   84.1   $\pm$    4.2   & \\ 
5071.61757  &   91.7   $\pm$    5.8   & \\ 
5071.62478  &  100.7   $\pm$    2.6   & \\ 
5071.63199  &  93.9    $\pm$    3.2    & \\
\noalign{\smallskip}
\hline
\noalign{\smallskip}
5144.55811  &    101.0  $\pm$  10.1 &  EFOSC2 \\
5144.59234  &     97.1  $\pm$  10.1 &  \\
5144.63989  &     76.5  $\pm$  12.6 &  \\
5145.52594  & -28.5 $\pm$  9.9  & \\
5145.61524  & -16.9 $\pm$  7.8 &  \\
5145.65247  &  -7.1 $\pm$ 10.2 &  \\
5146.54220  &   2.5 $\pm$  8.0 &  \\
5146.64161  & -27.0 $\pm$  8.1 &  \\ 
5146.69219  & -57.2 $\pm$  8.9 &  \\
5147.59722  &  82.4  $\pm$ 10.4  &  \\
5147.64906 &   81.5  $\pm$ 9.5  &  \\
5147.66103  &  85.2  $\pm$ 7.2  &  \\
5147.67897  &  80.0  $\pm$ 9.8  &  \\    
\noalign{\smallskip}
\hline

\end{tabular}
\end{center}
\end{table}

\begin{table}
\caption{J1138$-$0035}
\label{RV2}
\begin{center}
\begin{tabular}{lrl}
\hline
\noalign{\smallskip}
mid$-$HJD & RV [${\rm km\,s^{-1}}$] & Instrument\\
$-2\,450\,000$ & & \\
\noalign{\smallskip}
\hline
\noalign{\smallskip}
1629.831447 &   -16.0 $\pm$   33.7 &  SDSS  \\ 
1629.861748 & -131.2  $\pm$    5.8    & \\                                       
1629.875990 &  -139.6 $\pm$    10.5   & \\                                        
1629.890197 &  -124.6 $\pm$    11.9   & \\                                        
1630.849549 &    42.3 $\pm$    12.3   & \\                                        
1630.861782 &    -8.6 $\pm$     5.8   & \\                                        
1658.666991 &    16.0 $\pm$     6.1   & \\                                        
1658.723102 &  -127.9 $\pm$     2.4   & \\                                        
1658.737309 &  -120.4 $\pm$     7.4   & \\                                        
1658.749439 &   -81.3 $\pm$    12.2   & \\
\noalign{\smallskip}
\hline
\noalign{\smallskip}
2720.30421 &   -73.0 $\pm$    32.8   &  SAAO \\
2720.31674 &    -9.9 $\pm$    12.2   & \\
3101.42483 &   166.9 $\pm$    29.9   & \\
3101.44590 &    26.1 $\pm$    19.6   & \\
\noalign{\smallskip}
\hline
\noalign{\smallskip}
4186.50643  & -146.0  $\pm$   5.5  &  IDS \\
4186.52047  & -130.5  $\pm$   6.5  & \\ 
4187.51138  &  -76.5  $\pm$   6.8  & \\  
4187.52540  & -133.0  $\pm$   6.8  & \\ 
4188.55849  & -115.9  $\pm$   5.6  & \\ 
4188.57251  & -135.0  $\pm$   6.5  & \\ 
4189.46081  &    4.8  $\pm$   5.2  & \\ 
4189.47485  &   69.2  $\pm$   5.2  & \\ 
4190.52859  &  117.1  $\pm$   6.0  & \\ 
4190.54262  &  165.7  $\pm$   5.1  & \\ 
\noalign{\smallskip}
\hline
\noalign{\smallskip}
4991.40638  &     149.6 $\pm$  18.6   & IDS \\
4991.40999  &     143.0 $\pm$  15.0   & \\
4991.41360  &     151.1 $\pm$  15.8  & \\
4991.41722  &     148.4 $\pm$  15.0  & \\
4991.42083  &     163.3 $\pm$  18.2  & \\
4991.42444  &     162.0 $\pm$  18.4  & \\
4991.42964  &     182.6 $\pm$  15.9  & \\
4991.43325  &     172.7 $\pm$  21.8  & \\
4991.43687  &     192.4 $\pm$  17.6  & \\
4991.44048  &     193.8 $\pm$  18.3  & \\
4991.44409  &     156.9 $\pm$  18.8  & \\
4991.44770  &     169.9 $\pm$  15.0  & \\
4991.45131  &     151.1 $\pm$  17.2  & \\
4991.45492  &     151.7 $\pm$  17.4  & \\
\noalign{\smallskip}
\hline
\noalign{\smallskip}
5240.64268  &  38.5 $\pm$ 8.5   & Goodman \\
5240.64678  &  59.9 $\pm$ 7.0   & \\
5240.65068  &  59.2 $\pm$ 5.1   & \\
5240.65448  & 108.3 $\pm$ 8.6   & \\
5240.65828  & 116.8 $\pm$ 3.6   & \\
5240.75829  & -13.8 $\pm$ 5.3   & \\
5240.76549  & -49.6 $\pm$ 5.5   & \\
\noalign{\smallskip}
\hline

\end{tabular}
\end{center}
\end{table}

\begin{table}
\caption{J1505+1108}
\label{RV3}
\begin{center}
\begin{tabular}{lrl}
\hline
\noalign{\smallskip}
mid$-$HJD & RV [${\rm km\,s^{-1}}$] & Instrument\\
$-2\,450\,000$ & & \\
\noalign{\smallskip}
\hline
\noalign{\smallskip}
3848.858414 &      5.3  $\pm$  12.0 &  SDSS \\
3848.906794 &    27.7   $\pm$  12.2 & \\
3849.863669 &   -126.6  $\pm$  10.0 & \\
3850.893113 &    -96.1  $\pm$   8.1 & \\
\noalign{\smallskip}
\hline
\noalign{\smallskip}
4600.479632 &   -48.0 $\pm$     8.0 & TWIN \\
4600.487332 &   -65.4 $\pm$    10.0 & \\
4692.348022 &    22.2 $\pm$    13.0 & \\
4694.404665 &   -44.7 $\pm$     9.7 & \\
4696.399472 &  -148.2 $\pm$    10.8 & \\
4980.44927  &  -130.0 $\pm$      9.2 & \\
4980.53561  &  -175.5 $\pm$     9.9 & \\
4981.57813  &   -34.5 $\pm$     13.8 & \\
4982.55136  &     5.4 $\pm$     12.1 & \\
\noalign{\smallskip}
\hline
\noalign{\smallskip}
4936.672663 &    -37.3 $\pm$   2.8  & GMOS \\
4936.676403 &    -31.0 $\pm$   3.0 & \\
4936.680142 &    -34.7 $\pm$   3.0 & \\
4936.683881 &    -32.5 $\pm$   3.1 & \\ 
4936.871029 &     -3.5 $\pm$   2.8 & \\ 
4936.874768 &     -2.1 $\pm$   2.8 & \\ 
4936.878512 &      0.9 $\pm$   2.8 & \\ 
4936.882253 &     -5.5  $\pm$  2.9 & \\ 
4937.64103  &      0.8 $\pm$   3.1 & \\  
4937.64477  &      7.0  $\pm$  2.9 & \\ 
4937.64852  &      6.1 $\pm$   3.0 & \\ 
4937.65226  &      4.4 $\pm$   2.9 & \\     
4937.85083  &   -158.7 $\pm$   2.9 & \\ 
4937.85457  &   -161.9 $\pm$   2.9 & \\  
4937.85831  &   -158.5 $\pm$   2.9 & \\ 
4937.86206  &   -158.7  $\pm$  3.0 & \\ 
4938.75627  &   -150.6  $\pm$  2.5 & \\ 
4938.76001  &   -149.4 $\pm$   2.5 & \\ 
4938.76375  &   -156.3 $\pm$   2.5 & \\ 
4938.76749  &   -146.0 $\pm$   2.5 & \\ 
4939.67990  &    -31.1 $\pm$   2.8 & \\ 
4939.68364  &    -24.8 $\pm$   2.9 & \\ 
4939.68738  &    -26.7 $\pm$   2.8 & \\     
4939.69112  &    -23.6 $\pm$   2.8 & \\ 
4943.63233  &     -2.9 $\pm$   3.0 & \\ 
4943.63607  &     -9.2 $\pm$   2.9 & \\ 
4943.63981  &     -7.8 $\pm$   2.9 & \\ 
4943.64355  &    -10.3 $\pm$   2.7 & \\ 
\noalign{\smallskip}
\hline

\end{tabular}
\end{center}
\end{table}

\begin{table}
\caption{J1654+3037}
\label{RV4}
\begin{center}
\begin{tabular}{lrl}
\hline
\noalign{\smallskip}
mid$-$HJD & RV [${\rm km\,s^{-1}}$] & Instrument\\
$-2\,450\,000$ & & \\
\noalign{\smallskip}
\hline
\noalign{\smallskip}
2789.917095 &   119.4  $\pm$   9.7 &  SDSS \\
2789.933032 &    84.5  $\pm$   7.7  & \\
2790.913235 &   146.7  $\pm$   7.2  & \\
2790.929502 &   130.3  $\pm$   8.5  & \\
\noalign{\smallskip}
\hline
\noalign{\smallskip}
4586.567656 &   -6.0  $\pm$   7.8   & TWIN \\
4692.367579 &   155.2 $\pm$    8.0  & \\
4693.380826 &   148.1 $\pm$    8.1  & \\
4694.433521 &   155.9 $\pm$    8.0   & \\
5037.44285  &   70.1 $\pm$ 10.6  & \\
5037.47019  &  150.9 $\pm$  8.4  & \\
5037.50213  &  156.2 $\pm$ 11.1  & \\
5038.42271  &  -21.8 $\pm$ 11.0  & \\
5038.48616  &  155.0 $\pm$ 10.8  & \\
5038.49857  &  161.0 $\pm$  7.9  & \\
\noalign{\smallskip}
\hline
\noalign{\smallskip}
4988.47623  &    -4.3 $\pm$ 25.5 & IDS \\
4988.49036  &    56.8 $\pm$ 16.6 & \\
4988.50437  &    97.9 $\pm$ 17.5 & \\
4988.52125  &   129.9 $\pm$ 15.6 & \\
4988.53530  &   172.6 $\pm$ 17.3 & \\
4988.54942  &   179.7 $\pm$ 18.0 & \\
4988.56430  &   150.0 $\pm$ 16.0 & \\
4991.47179  &   -84.8 $\pm$ 17.7 & \\
4991.47888  &   -82.0 $\pm$ 17.1 & \\
4991.48596  &   -92.2 $\pm$ 16.6 & \\
4991.49305  &   -58.5 $\pm$  7.1 & \\
4991.50013  &   -62.6 $\pm$ 17.5 & \\
4991.50927  &   -44.1 $\pm$ 16.5 & \\
4991.51753  &   -11.2 $\pm$ 15.1 & \\
4991.52577  &    23.0 $\pm$ 18.4 & \\
4991.53402  &    33.3 $\pm$ 16.2 & \\
4991.54224  &    66.3 $\pm$ 16.0 & \\
4991.55155  &   103.8 $\pm$ 30.1 & \\
4991.55981  &   128.0 $\pm$ 17.6 & \\
4991.56805  &   142.9 $\pm$ 16.5 & \\
4991.57629  &   161.2 $\pm$ 17.1 & \\
4991.58453  &   178.5 $\pm$ 16.4 & \\
\noalign{\smallskip}
\hline

\end{tabular}
\end{center}
\end{table}

\begin{table}
\caption{J1726+2744}
\label{RV5}
\begin{center}
\begin{tabular}{lrl}
\hline
\noalign{\smallskip}
mid$-$HJD & RV [${\rm km\,s^{-1}}$] & Instrument\\
$-2\,450\,000$ & & \\
\noalign{\smallskip}
\hline
\noalign{\smallskip}
3905.819525 &   -168.9 $\pm$    10.9 & SDSS \\
3905.833513 &   -157.3 $\pm$    10.7 & \\
3905.853553 &   -133.2 $\pm$    10.5 & \\
3905.866007 &   -123.5 $\pm$    15.8 & \\
\noalign{\smallskip}
\hline
\noalign{\smallskip}
4979.59252  &   -125.2  $\pm$   14.4 & TWIN \\
4979.63382  &    -98.5  $\pm$   11.6 & \\
4980.47951  &   -155.4  $\pm$   12.9 & \\
4980.58604  &   -141.2  $\pm$   11.4 & \\
4981.53132  &   -120.0  $\pm$    8.0 & \\
4981.58892  &   -112.5  $\pm$   12.9 & \\
4981.62939  &    -46.7  $\pm$   12.0 & \\
4981.64123  &    -53.1  $\pm$    9.7 & \\
4981.65164  &    -35.2  $\pm$   16.9 & \\
4982.56453  &   -154.2  $\pm$   11.2 & \\
4982.65151  &    -65.0  $\pm$   21.6 & \\
4983.53806  &   -152.7  $\pm$   11.5 & \\
4983.54720  &   -176.7  $\pm$   12.0 & \\
4983.55638  &   -150.5  $\pm$   15.2 & \\
4983.56557  &   -146.3  $\pm$   24.3 & \\
4983.57468  &   -124.9  $\pm$   13.8 & \\
5037.45679  &     39.0  $\pm$  12.0 & \\
5037.48545  &     62.0  $\pm$  12.7 & \\
5038.43771  &     38.0  $\pm$  16.1 & \\
5038.45381  &     58.0  $\pm$  12.9 & \\
5038.47045  &     80.0  $\pm$  13.0 & \\
5039.49762  &     69.0  $\pm$  12.7 & \\
5039.52128  &     87.0  $\pm$  12.0 & \\
5039.53838  &     85.0  $\pm$  12.9 & \\
\noalign{\smallskip}
\hline
\noalign{\smallskip}
4992.48445  &    -99.5  $\pm$  23.1 & IDS \\
4992.49617  &   -114.1  $\pm$  18.5 & \\
4992.50788  &    -81.2 $\pm$   13.7 & \\
\noalign{\smallskip}
\hline
\noalign{\smallskip}
5068.47186  &  -92.7  $\pm$    5.2 & ISIS \\
5068.47906  &  -79.4  $\pm$    4.4   & \\
5068.48628  &  -73.9  $\pm$    5.3   & \\
5069.45355  & -120.3  $\pm$    3.2   & \\
5069.46426  & -109.9  $\pm$    2.7   & \\
5069.46786  &  -98.2  $\pm$    4.5   & \\
5069.47855  &  -79.9  $\pm$    3.8   & \\
\noalign{\smallskip}
\hline
\end{tabular}
\end{center}
\end{table}

\begin{table}
\caption{J2046-0454}
\label{RV6}
\begin{center}
\begin{tabular}{lrl}
\hline
\noalign{\smallskip}
mid$-$HJD & RV [${\rm km\,s^{-1}}$] & Instrument\\
$-2\,450\,000$ & & \\
\noalign{\smallskip}
\hline
\noalign{\smallskip}
3269.661429  &   109.6 $\pm$   13.3 &  SDSS \\
3269.675556  &   128.1 $\pm$   9.6 & \\
3269.691435  &   179.6 $\pm$   8.8 & \\
\noalign{\smallskip}
\hline
\noalign{\smallskip}
4645.79103   &   181.6  $\pm$   5.0 & FORS1 \\
4645.79259  &    185.0  $\pm$   3.1 & \\
\noalign{\smallskip}
\hline
\noalign{\smallskip}
4692.51274  &  28.4  $\pm$  13.4  & TWIN \\
4692.52696  &  30.7  $\pm$  14.3  & \\
4693.42472  & 227.7  $\pm$  11.9  & \\
4693.47199  & 137.2  $\pm$  20.6  & \\
4696.49294   & 42.0  $\pm$   9.3  & \\
4696.54469  & 177.8  $\pm$  10.9  & \\
4696.60171  & 171.6  $\pm$   9.8  & \\
4979.61251  &      90.6  $\pm$   6.8   & \\ 
4979.65127  &     -32.8  $\pm$  21.3   & \\
5035.46989  &     207.0  $\pm$  15.3   & \\ 
5035.49811  &     210.0  $\pm$  14.3   & \\
5036.50301  &      83.0  $\pm$  15.5   & \\
5037.52184  &     -25.0  $\pm$  10.8   & \\
5037.59514  &      16.0  $\pm$  15.0   & \\
\noalign{\smallskip}
\hline
\noalign{\smallskip}
4758.55029  & 206.1  $\pm$  22.5 & EFOSC2 \\
4758.55416  & 202.4  $\pm$  25.3 & \\
4758.55803  & 212.2  $\pm$  22.5 & \\
\noalign{\smallskip}
\hline
\end{tabular}
\end{center}
\end{table}

\begin{table}
\caption{J2256+0656}
\label{RV7}
\begin{center}
\begin{tabular}{lrl}
\hline
\noalign{\smallskip}
mid$-$HJD & RV [${\rm km\,s^{-1}}$] & Instrument\\
$-2\,450\,000$ & & \\
\noalign{\smallskip}
\hline
\noalign{\smallskip}
3710.557488  &    11.8  $\pm$ 10.5 & SDSS \\
3710.571464  &     1.1  $\pm$   8.6 & \\
3710.588935  &   -15.5  $\pm$   9.7 & \\
\noalign{\smallskip}
\hline
\noalign{\smallskip}
4694.610760  &    78.0  $\pm$   6.0 &  TWIN \\
4694.676593  &    79.3  $\pm$   1.0  & \\
4694.689883  &    87.1  $\pm$  6.0  & \\
5035.48389   &  -140.9  $\pm$  23.8  & \\
5035.50848   &   -96.8  $\pm$  15.6  & \\
5036.51782   &    95.0  $\pm$  15.0  & \\
5036.61657   &    29.9  $\pm$  10.7  & \\
5036.56824   &    81.7  $\pm$  11.2  & \\
5037.53430   &  -129.6  $\pm$  17.9  & \\
5037.54784   &  -120.8  $\pm$  11.1  & \\
5037.55965   &  -134.9  $\pm$  11.0  & \\
5037.60833   &  -116.5  $\pm$  11.0  & \\
5038.56302   &     96.6 $\pm$   10.6  & \\
5038.57506   &   115.1  $\pm$  12.7  & \\
5038.62975   &   104.1  $\pm$  15.3  & \\
5038.64304   &   105.0  $\pm$  11.8  & \\
5038.65573   &   104.1  $\pm$  13.1  & \\
5039.64785   &  -123.9  $\pm$  12.5  & \\
\noalign{\smallskip}
\hline
\noalign{\smallskip}
5048.981437  &  16.4  $\pm$   4.2 & GMOS \\
5049.002417  &  28.5  $\pm$   4.3 & \\
5049.023396  &  43.5  $\pm$   4.2 & \\
5049.044376  &  51.0  $\pm$   4.3 & \\
5077.83332   &  129.0  $\pm$  5.6 & \\
5077.85483   &  116.5  $\pm$  5.3 & \\
\noalign{\smallskip}
\hline
\noalign{\smallskip}
5068.53076   & -36.5  $\pm$     3.4 & ISIS \\
5068.53111   & -30.2  $\pm$     3.5     & \\
5068.53826   & -26.5  $\pm$     2.3     & \\
5068.54815   & -10.3  $\pm$     2.7     & \\
5068.55536   & -12.8  $\pm$     3.2     & \\
5068.56257   &  -1.5  $\pm$     2.3     & \\
5069.53785   &  41.6  $\pm$     3.7     & \\
5069.54507   &  48.0   $\pm$    3.8     & \\
5069.55228   &  45.8   $\pm$    2.3     & \\
5069.56913   &  48.2   $\pm$    3.6     & \\
5069.57635   &  26.9  $\pm$     3.8    & \\
5069.58358   &  18.2  $\pm$     3.8     & \\
5070.62493   & -48.4  $\pm$     4.0     & \\
5070.63214   & -35.8   $\pm$    4.0     & \\
5070.63942   & -21.1   $\pm$    4.1     & \\
5070.65603   & -20.0   $\pm$    2.6     & \\
5070.66324   &  -1.5  $\pm$     4.5     & \\
5070.67045   &   2.6  $\pm$     4.6     & \\
5071.39970   &   8.8    $\pm$   4.8     & \\
5071.39997   &  21.3  $\pm$     4.2     & \\
5071.40718   & 33.8   $\pm$     3.6    & \\
5071.42392   &  50.5  $\pm$     3.8     & \\
5071.43114   &  50.8  $\pm$     3.9    & \\
\noalign{\smallskip}
\hline
\end{tabular}
\end{center}
\end{table}

\end{appendix}

\end{document}